\documentclass[fleqn,usenatbib]{mnras}

\usepackage{newtxtext,newtxmath}

\usepackage[T1]{fontenc}
\usepackage{ae,aecompl}
\usepackage{booktabs,subcaption,amsfonts}
\usepackage{multirow}

\newcommand{\NGTS}{NGTS}

\newcommand{\ariadne}{\textsc{ariadne}\xspace}
\newcommand{\bruce}{\textsc{bruce}\xspace}
\newcommand{\allesfitter}{\textsc{allesfitter}\xspace}
\newcommand{\ispec}{\textsc{iSpec}\xspace}
\newcommand{\ellc}{\texttt{ellc}\xspace}
\newcommand{\dynesty}{\texttt{dynesty}\xspace}
\newcommand{\emcee}{\texttt{emcee}\xspace}
\newcommand{\celerite}{\texttt{celerite}\xspace}

\newcommand{\orion}{\textsc{orion}\xspace}

\newcommand{\kms}{km\,s$^{-1}$}
\newcommand{\ms}{m\,s$^{-1}$}
\newcommand{\masy}{mas\,y$^{-1}$}
\newcommand{\mpl}{\mbox{$M_{p}$}}
\newcommand{\rpl}{\mbox{$R_{p}$}}
\newcommand{\mstar}{\mbox{$M_{s}$}}
\newcommand{\rstar}{\mbox{$R_{s}$}}
\newcommand{\mjup}{\mbox{$M_{J}$}}
\newcommand{\rjup}{\mbox{$R_{J}$}}
\newcommand{\msun}{\mbox{$M_{\odot}$}\xspace}
\newcommand{\rsun}{\mbox{$R_{\odot}$}\xspace}

\newcommand{\gccc}{g\,cm$^{-3}$}

\newcommand{\teff}{$T_{\rm eff}$\xspace}
\newcommand{\logg}{$\log g$}
\newcommand{\feh}{\mbox{[Fe/H]}\xspace}
\newcommand{\vsini}{$v\sin{i}$}
\newcommand{\inflationflux}{$2 \times 10^{5}$ Wm$^{-2}$}

\newcommand{\starA}{NGTS-15}
\newcommand{\Akamp}{\mbox{$106 \pm 7$}}%Note units m/s.
\newcommand{\Agamma}{\mbox{$34.68 \pm 0.04 $}}%Note units km/s
\newcommand{\ARA}{\mbox{$04^{\rmn{h}} 53^{\rmn{m}} 25\fs27$}} %2MASS
\newcommand{\ADec}{\mbox{$-32\degr 48\arcmin 01\farcs 25$}} %2MASS
\newcommand{\ApropRA}{\mbox{$5.687\pm0.027$}} %GAIADR2 
\newcommand{\ApropDec}{\mbox{$1.316\pm0.035$}} %GAIADR2
\newcommand{\Astarmass}{\mbox{$0.995\pm0.015\pm0.020$}}
\newcommand{\Astarradius}{\mbox{$0.954 \pm 0.023\pm0.079$}}

\newcommand{\Ateff}{\mbox{$5600\pm150$}}%SED
\newcommand{\Ametal}{\mbox{$0.15\pm0.1$}}
\newcommand{\Adist}{\mbox{$791\pm19\pm59$}}%SED (pc)
\newcommand{\Alogg}{\mbox{$4.40\,^{+0.05}_{-0.06}\pm0.11$}}

\newcommand{\Agalu}{\mbox{$26.875 \pm 0.017$}}
\newcommand{\Agalv}{\mbox{$-10.106 \pm 0.024$}}
\newcommand{\Agalw}{\mbox{$-14.065 \pm 0.023$}}
\newcommand{\Avsini}{$<2$}
\newcommand{\Aage}{$3.28\,^{+1.06}_{-1.38}\,^{+4.10}_{-3.28}$} 
\newcommand{\AVmag}{$14.562 \pm 0.029$}
\newcommand{\ABmag}{$15.294 \pm 0.040$}

\newcommand{\Agmag}{$14.890 \pm 0.034$}
\newcommand{\Armag}{$14.341 \pm 0.064$}
\newcommand{\Aimag}{$14.219 \pm 0.094$}
\newcommand{\AGAIAmag}{$14.4299 \pm 0.0003$}

\newcommand{\AJmag}{$13.341\pm0.030$}
\newcommand{\AHmag}{$12.967\pm0.035$}
\newcommand{\AKmag}{$12.907\pm0.036$}
\newcommand{\AWmag}{$12.824 \pm 0.024$}
\newcommand{\AWWmag}{$12.863 \pm 0.025$}
\newcommand{\ATESSmag}{$13.964 \pm 0.006$}
\newcommand{\Aspectype}{G6V}
\newcommand{\Aparallax}{$1.242 \pm 0.017$}
\newcommand{\AAv}{$0.025 \pm 0.008 \pm 0.018$}

\newcommand{\planetA}{NGTS-15b}
\newcommand{\Abperiod}{\mbox{$3.27623 \pm 0.00001$}}

\newcommand{\Abduration}{\mbox{$2.504\,^{+0.060}_{-0.058}$}}%
\newcommand{\Abtc}{\mbox{$2458405.0558 \pm 0.0007$}}
\newcommand{\Abmass}{\mbox{$0.751\,^{+0.102}_{-0.088}$}}%
\newcommand{\Abradius}{\mbox{$1.10\pm0.10$}}%
\newcommand{\Abdensity}{\mbox{$0.74\,^{+0.11}_{-0.10}$}}%errors??
\newcommand{\AbTeq}{\mbox{$1146\pm47$}}%
\newcommand{\Abflux}{\mbox{($5.78\pm1.33) \times 10^{5}$}}%
\newcommand{\Abrratio}{\mbox{$0.119\pm0.003$}}
\newcommand{\Abau}{\mbox{$0.0441 \pm 0.0046$}}%
\newcommand{\Abaoverr}{\mbox{$10.00\pm0.59$}}%

\newcommand{\Abimpact}{\mbox{$0.509\,^{+0.086}_{-0.112}$}}%

\newcommand{\ANGTSdil}{$0.04\,^{+0.01}_{-0.02}$}

\newcommand{\starB}{NGTS-16}
\newcommand{\Bkamp}{\mbox{$74\,^{+14}_{-12}$}}%Note units m/s.
\newcommand{\Bgamma}{\mbox{$29.13 \pm 0.01 $}}%Note units km/s
\newcommand{\BRA}{\mbox{$03^{\rmn{h}} 53^{\rmn{m}} 03\fs34$}} %2MASS
\newcommand{\BDec}{\mbox{$-30\degr 48\arcmin 16\farcs 71$}} %2MASS
\newcommand{\BpropRA}{\mbox{$9.772\pm0.031$}} %UCAC4 
\newcommand{\BpropDec}{\mbox{$-4.673\pm0.043$}} %UCAC4
\newcommand{\Bstarmass}{\mbox{$1.002\pm0.011\pm0.029$}}
\newcommand{\Bstarradius}{\mbox{$1.213\,^{+0.040}_{-0.032}\pm0.102$}}

\newcommand{\Bteff}{\mbox{$5550\pm150$}} %ARIADNE + iSpec errors
\newcommand{\Bmetal}{\mbox{$0.35\pm0.1$}}
\newcommand{\Bdist}{\mbox{$892\,^{+29}_{-23}\pm63$}}%ARIADNE (errors too low?) (pc)
\newcommand{\Blogg}{\mbox{$4.35\,^{+0.11}_{-0.08}\pm0.14$}} %ARIADNE

\newcommand{\Bgalu}{\mbox{$23.193 \pm 0.004$}}
\newcommand{\Bgalv}{\mbox{$-1.914 \pm 0.004$}}
\newcommand{\Bgalw}{\mbox{$-15.158 \pm 0.007$}}
\newcommand{\Bvsini}{$2.5 \pm 0.8$}
\newcommand{\Bage}{$10.29\,^{+0.74}_{-0.53}\pm1.38$} 
\newcommand{\BVmag}{$14.364 \pm 0.039$}
\newcommand{\BBmag}{$15.140 \pm 0.006$}

\newcommand{\Bgmag}{$14.743 \pm 0.012$}
\newcommand{\Brmag}{$14.148 \pm 0.018$}
\newcommand{\Bimag}{$13.960 \pm 0.078$}
\newcommand{\BGAIAmag}{$14.2311 \pm 0.0002$}

\newcommand{\BJmag}{$13.090\pm0.024$}
\newcommand{\BHmag}{$12.735\pm0.027$}
\newcommand{\BKmag}{$12.647\pm0.033$}
\newcommand{\BWmag}{$12.638 \pm 0.025$}
\newcommand{\BWWmag}{$12.710 \pm 0.025$}
\newcommand{\BTESSmag}{$13.743 \pm 0.007$}
\newcommand{\Bspectype}{G7V}
\newcommand{\Bparallax}{$1.084 \pm 0.021$}
\newcommand{\BAv}{$0.013 \pm 0.004 \pm 0.011$}

\newcommand{\planetB}{NGTS-16b}
\newcommand{\Bbperiod}{\mbox{$4.84532 \pm 0.00002$}}

\newcommand{\Bbduration}{\mbox{$3.061\,^{+0.143}_{-0.136}$}}% "total transit duration"
\newcommand{\Bbtc}{\mbox{$2458435.6054 \pm 0.0013$}}
\newcommand{\Bbmass}{\mbox{$0.667\,^{+0.157}_{-0.129}$}}%
\newcommand{\Bbradius}{\mbox{$1.30\,^{+0.13}_{-0.12}$}}%
\newcommand{\Bbdensity}{\mbox{$0.38\,^{+0.19}_{-0.12}$}}%
\newcommand{\BbTeq}{\mbox{$1177 \pm 59$}}%
\newcommand{\Bbflux}{\mbox{($6.60\pm1.56) \times 10^{5}$}}%
\newcommand{\Bbrratio}{\mbox{$0.110\pm0.005$}}
\newcommand{\Bbau}{\mbox{$0.0523 \pm 0.0064$}}%
\newcommand{\Bbaoverr}{\mbox{$9.29\pm0.83$}}%

\newcommand{\Bbimpact}{\mbox{$0.807\,^{+0.049}_{0.036}$}}%

\newcommand{\BTESSdil}{$0.11\,^{+0.03}_{-0.05}$}

\newcommand{\starC}{NGTS-17}
\newcommand{\Ckamp}{\mbox{$93\,^{+20}_{-17}$}}%Note units m/s.
\newcommand{\Cgamma}{\mbox{$34.81 \pm 0.07 $}}%Note units km/s
\newcommand{\CRA}{\mbox{$04^{\rmn{h}} 51^{\rmn{m}} 36\fs14$}} %2MASS
\newcommand{\CDec}{\mbox{$-34\degr 13\arcmin 34\farcs 37$}} %2MASS
\newcommand{\CpropRA}{\mbox{$-0.190\pm0.031$}} %UCAC4 
\newcommand{\CpropDec}{\mbox{$-11.812\pm0.034$}} %UCAC4
\newcommand{\Cstarmass}{\mbox{$1.025\,^{+0.015}_{-0.014}\pm0.03$}}
\newcommand{\Cstarradius}{\mbox{$1.337\pm0.038\pm0.119$}}

\newcommand{\Cteff}{\mbox{$5650\pm100$}} % ARIADNE + iSpec errors
\newcommand{\Cmetal}{\mbox{$0.15\pm0.1$}}
\newcommand{\Cdist}{\mbox{$1047\,^{29}_{-27}\pm108$}} %ARIADNE (pc)
\newcommand{\Clogg}{\mbox{$4.00\pm0.09\pm0.08$}}
\newcommand{\Cgalu}{\mbox{$26.073 \pm 0.031$}}
\newcommand{\Cgalv}{\mbox{$-10.465 \pm 0.046$}}
\newcommand{\Cgalw}{\mbox{$-14.478 \pm 0.044$}}
\newcommand{\Cvsini}{$4.1 \pm 1.0$} %iSpec
\newcommand{\Cparallax}{$0.932 \pm 0.020$}

\newcommand{\Cage}{$9.22\,^{+0.49}_{-0.48}\pm1.32$}
\newcommand{\CVmag}{$14.326 \pm 0.030$}
\newcommand{\CBmag}{$15.043 \pm 0.033$}

\newcommand{\Cgmag}{$14.626 \pm 0.013$}
\newcommand{\Crmag}{$14.161 \pm 0.043$}
\newcommand{\Cimag}{$13.998 \pm 0.081$}
\newcommand{\CGAIAmag}{$14.2136 \pm 0.0004$}

\newcommand{\CJmag}{$13.114\pm0.026$}
\newcommand{\CHmag}{$12.867\pm0.027$}
\newcommand{\CKmag}{$12.724\pm0.029$}
\newcommand{\CWmag}{$12.739 \pm 0.023$}
\newcommand{\CWWmag}{$12.747 \pm 0.027$}
\newcommand{\CTESSmag}{$13.767 \pm 0.006$}
\newcommand{\Cspectype}{G4V}
\newcommand{\CAv}{$0.028 \pm 0.008 \pm 0.019$}

\newcommand{\planetC}{NGTS-17b}
\newcommand{\Cbperiod}{\mbox{$3.24253 \pm 0.00001$}}

\newcommand{\Cbduration}{\mbox{$3.391\,^{+0.076}_{-0.073}$}}%
\newcommand{\Cbtc}{\mbox{$2458442.5219 \pm 0.0009$}}
\newcommand{\Cbmass}{\mbox{$0.764\,^{+0.195}_{-0.164}$}}%
\newcommand{\Cbradius}{\mbox{$1.24 \pm 0.11$}}%
\newcommand{\Cbdensity}{\mbox{$0.50\,^{+0.27}_{-0.17}$}}%
\newcommand{\CbTeq}{\mbox{$1457 \pm 50$}}%
\newcommand{\Cbflux}{\mbox{($1.58\pm0.34) \times 10^{6}$}}%
\newcommand{\Cbrratio}{\mbox{$0.095\pm0.001$}}
\newcommand{\Cbau}{\mbox{$0.0391\pm0.0043$}}%
\newcommand{\Cbaoverr}{\mbox{$6.28\pm0.40$}}%

\newcommand{\Cbimpact}{\mbox{$0.688\,^{+0.043}_{-0.056}$}}%

\newcommand{\CTESSdil}{$0.09\,^{+0.01}_{-0.02}$}

\newcommand{\starD}{NGTS-18}
\newcommand{\Dkamp}{\mbox{$62 \pm 5$}}%Note units m/s.
\newcommand{\Dgamma}{\mbox{$5.19 \pm 0.02 $}}%Note units km/s
\newcommand{\DRA}{\mbox{$12^{\rmn{h}} 02^{\rmn{m}} 11 \fs09$}} %2MASS
\newcommand{\DDec}{\mbox{$-35\degr 32\arcmin 54\farcs 99$}} %2MASS
\newcommand{\DpropRA}{\mbox{$-2.416\pm0.040 $}} %UCAC4 
\newcommand{\DpropDec}{\mbox{$1.487\pm0.025$}} %UCAC4
\newcommand{\Dstarmass}{\mbox{$1.003\,^{+0.020}_{-0.012}\pm0.022$}}
\newcommand{\Dstarradius}{\mbox{$1.392\,^{+0.057}_{-0.058}\pm0.201$}}

\newcommand{\Dteff}{\mbox{$5610\pm150$}}%SED
\newcommand{\Dmetal}{\mbox{$0.15\pm0.1$}}
\newcommand{\Ddist}{\mbox{$1108\,^{+44}_{-47}\pm158$}}%SED (pc)
\newcommand{\Dlogg}{\mbox{$4.16 \pm 0.05 \pm 0.05$}}
\newcommand{\Dgalu}{\mbox{$9.385 \pm 0.008$}}
\newcommand{\Dgalv}{\mbox{$7.917 \pm 0.020$}}
\newcommand{\Dgalw}{\mbox{$9.548 \pm 0.011$}}
\newcommand{\Dvsini}{$<0.5$}
\newcommand{\Dparallax}{$0.884 \pm 0.034$}

\newcommand{\Dage}{$10.84\,^{+0.40}_{-0.78}\pm1.22$}
\newcommand{\DVmag}{$14.540 \pm 0.038$}
\newcommand{\DBmag}{$15.322 \pm 0.046$}

\newcommand{\Dgmag}{$14.871 \pm 0.021$}
\newcommand{\Drmag}{$14.387 \pm 0.080$}
\newcommand{\Dimag}{$14.189 \pm 0.122$}
\newcommand{\DGAIAmag}{$14.3896 \pm 0.0003$}

\newcommand{\DJmag}{$13.265 \pm 0.030$}
\newcommand{\DHmag}{$12.888 \pm 0.023$}
\newcommand{\DKmag}{$12.870 \pm 0.032$}
\newcommand{\DWmag}{$12.768 \pm 0.023$}
\newcommand{\DWWmag}{$12.832 \pm 0.026$}
\newcommand{\DTESSmag}{$13.907 \pm 0.006$}
\newcommand{\Dspectype}{G5V}
\newcommand{\DAv}{$0.11 \pm 0.028 \pm 0.022$}

\newcommand{\planetD}{NGTS-18b}
\newcommand{\Dbperiod}{\mbox{$3.05125 \pm 0.00001$}}

\newcommand{\Dbduration}{\mbox{$3.601\,^{+0.055}_{-0.047}$}}%
\newcommand{\Dbtc}{\mbox{$2458564.4506 \pm 0.0007$}}
\newcommand{\Dbmass}{\mbox{$0.409\,^{+0.081}_{-0.063}$}}%
\newcommand{\Dbradius}{\mbox{$1.21 \pm 0.18$}}%
\newcommand{\Dbdensity}{\mbox{$0.28\,^{+0.25}_{-0.12}$}}%
\newcommand{\DbTeq}{\mbox{$1381\,^{+55}_{-53}$}}%
\newcommand{\Dbflux}{\mbox{($1.15\pm0.37) \times 10^{6}$}}%
\newcommand{\Dbrratio}{\mbox{$0.089\pm0.001$}}
\newcommand{\Dbau}{\mbox{$0.0448\pm0.0068$}}%
\newcommand{\Dbaoverr}{\mbox{$6.97\pm0.27$}}%

\newcommand{\Dbimpact}{\mbox{$0.183\,^{+0.147}_{-0.115}$}}%

\usepackage{graphicx}	% Including figure files
\usepackage{amsmath}	% Advanced maths commands
\usepackage{xspace}
\usepackage{multicol}
\usepackage[symbol]{footmisc}
\newcommand{\customfootnotetext}[2]{{% Group to localize change to footnote
  \renewcommand{\thefootnote}{#1}% Update footnote counter representation
  \footnotetext[0]{#2}}}

\title[NGTS 15b, 16b, 17b and 18b]{NGTS 15b, 16b, 17b and 18b: four hot Jupiters from the Next Generation Transit Survey}

\author[Rosanna H. Tilbrook et al.]{
\parbox{\textwidth}{
Rosanna H. Tilbrook,$^{1 \hyperlink{rht6@le.ac.uk}{\star}}$\vspace{0.1cm}
Matthew R. Burleigh,$^{1}$
Jean C. Costes,$^{2}$
Samuel Gill,$^{3,4}$
Louise Dyregaard Nielsen,$^{5}$
Jos{\'e} I. Vines,$^{6}$
Didier~Queloz,$^{7}$
Simon~T.~Hodgkin,$^{8}$
Hannah~L.~Worters,$^{9}$
Michael R.~Goad,$^{1}$
Jack~S.~Acton,$^{1}$
Beth A. Henderson,$^{1}$
David J. Armstrong,$^{3,4}$
David R. Anderson,$^{3,4}$
Daniel~Bayliss,$^{3,4}$
Fran\c{c}ois Bouchy,$^{5}$
Joshua~T.~Briegal,$^{7}$
Edward~M.~Bryant,$^{3,4}$
Sarah L. Casewell,$^{1}$
Alexander Chaushev,$^{10}$
Benjamin F. Cooke,$^{3,4}$
Philipp~Eigm\"uller,$^{11}$
Edward~Gillen,$^{12,7 \hyperlink{Ed}{\dagger}}$\vspace{0.05cm}
Maximilian~N.~G{\"u}nther,$^{13 \hyperlink{Max}{\ddagger}}$\vspace{0.05cm}
Aleisha Hogan,$^{1}$
James~S.~Jenkins,$^{6,14}$
Monika~Lendl,$^{5}$
James~McCormac,$^{3,4}$
Maximiliano~Moyano,$^{15}$
Liam Raynard,$^{1}$
Alexis~M.~S.~Smith,$^{11}$
St\'{e}phane~Udry,$^{5}$
Christopher~A.~Watson,$^{2}$
Richard~G.~West,$^{3,4}$
Peter~J.~Wheatley,$^{3,4}$
Hannes~Breytenbach,$^{9,16}$
Ramotholo~R.~Sefako,$^{9}$
Jessymol~K.~Thomas,$^{9}$
Douglas R. Alves$^{6}$
}
\\
\\
$^{1}$School of Physics and Astronomy, University of Leicester, LE1 7RH, UK\\
$^{2}$Astrophysics Research Centre, School of Mathematics and Physics, Queen's University Belfast, BT7 1NN Belfast, UK\\
$^{3}$Centre for Exoplanets and Habitability, University of Warwick, Gibbet Hill Road, Coventry CV4 7AL, UK\\
$^{4}$Dept.\ of Physics, University of Warwick, Gibbet Hill Road, Coventry CV4 7AL, UK\\
$^{5}$Observatoire de Gen{\`e}ve, Universit{\'e} de Gen{\`e}ve, 51 Ch. des Maillettes, 1290 Sauverny, Switzerland\\
$^{6}$Departamento de Astronomia, Universidad de Chile, Casilla 36-D, Santiago, Chile\\
$^{7}$Astrophysics Group, Cavendish Laboratory, J.J. Thomson Avenue, Cambridge CB3 0HE, UK\\
$^{8}$Institute of Astronomy, University of Cambridge, Madingley Rise, Cambridge CB3 0HA, UK\\
$^{9}$South African Astronomical Observatory, P.O Box 9, Observatory 7935, Cape Town, South Africa\\
$^{10}$Center for Astronomy and Astrophysics, TU Berlin, Hardenbergstr. 36, D-10623 Berlin, Germany\\
$^{11}$Institute of Planetary Research, German Aerospace Center, Rutherfordstrasse 2, 12489 Berlin, Germany\\
$^{12}$Astronomy Unit, Queen Mary University of London, Mile End Road, London E1 4NS, UK\\
$^{13}$Department of Physics, and Kavli Institute for Astrophysics and Space Research, Massachusetts Institute of Technology,\\
Cambridge, MA 02139, USA\\
$^{14}$Centro de Astrof\'isica y Tecnolog\'ias Afines (CATA), Casilla 36-D, Santiago, Chile.\\
$^{15}$Instituto de Astronom\'ia, Universidad Cat\'{o}lica del Norte, Angamos 0610, 1270709, Antofagasta, Chile\\
$^{16}$Department of Astronomy, University of Cape Town, Rondebosch 7700, Cape Town, South Africa
\vspace{-0.5cm}}
\date{\vspace{-0.3cm}Accepted XXX. Received YYY; in original form ZZZ}

\pubyear{2020}

\begin{document}
\definecolor{mygray}{gray}{0.6}
\label{firstpage}
\pagerange{\pageref{firstpage}--\pageref{lastpage}}
\maketitle

% Abstract of the paper
\begin{abstract}
    We report the discovery of four new hot Jupiters with the Next Generation Transit Survey (NGTS). \planetA, \planetB, \planetC, and \planetD\ are short-period ($P<5$d) planets orbiting G-type main sequence stars, with radii and masses between $1.10-1.30$ $R_J$ and $0.41-0.76$ $M_J$. By considering the host star luminosities and the planets' small orbital separations ($0.039-0.052$ AU), we find that all four hot Jupiters are highly irradiated and therefore occupy a region of parameter space in which planetary inflation mechanisms become effective. Comparison with statistical studies and a consideration of the planets' high incident fluxes reveals that \planetB, \planetC, and \planetD\ are indeed likely inflated, although some disparities arise upon analysis with current Bayesian inflationary models. However, the underlying relationships which govern radius inflation remain poorly understood. We postulate that the inclusion of additional hyperparameters to describe latent factors such as heavy element fraction, as well as the addition of an updated catalogue of hot Jupiters, would refine inflationary models, thus furthering our understanding of the physical processes which give rise to inflated planets.

\end{abstract}


%%%%%%%%%%%%%%%%%%%%%%%%%%%%%%%%%%%%%%%%%%%%%%%%%%

%%%%%%%%%%%%%%%%% BODY OF PAPER %%%%%%%%%%%%%%%%%%

\section{Introduction}

\customfootnotetext{\star}{\hypertarget{rht6@le.ac.uk}E-mail: \href{rht6@le.ac.uk}{rht6@le.ac.uk}}
\customfootnotetext{\dagger}{\hypertarget{Ed}{Winton Fellow}}
\customfootnotetext{\ddagger}{\hypertarget{Max}{Juan Carlos Torres Fellow}}

The field of exoplanet discovery has uncovered a cosmic zoo of planetary types which extends far beyond those of our Solar System. As some of the first exoplanets ever detected, a particularly immediate revelation was the existence of Jupiter-sized planets on extremely short orbits ($P<10$d) (e.g. \citealt{Mayor95, Charbonneau00, Henry00}). In spite of their apparent rarity, comprising only $<$1\% of systems \citep{Mayor11, Wright12, fressin13, Hsu2019}, these `hot Jupiters' are some of the most easily detectable exoplanets, as their large radii produce transits that are well above typical telescope noise limits, and their large masses and short orbital periods yield large radial velocity signals.

A distinctive yet poorly understood feature of the hot Jupiter population is the observation that many of these planets have radii that are larger than expected from theoretical models (e.g. \citealp{GuillotShowman02, Anderson11, Hartman12, Espinoza16, Raynard18}). Whilst this feature appears to correlate with incident flux \citep{Laughlin11, Weiss13, Thorngren18} this alone does not provide a sufficient explanation for inflation, and thus the underlying driving mechanisms continue to be debated (see, e.g., \citealp{Spiegel13}, for a comprehensive review). Indeed, evolutionary models that incorporate stellar irradiation and heavy metals can explain the unexpected radii of some hot Jupiters \citep{Fortney07, Baraffe08}, but are yet to adequately describe the high-irradiation regime.

Efforts to characterise the relationship between inflation and incident flux have nevertheless provided valuable insight into the topic. In particular, the physical processes that cause inflation are believed to become effective in planets that are irradiated at fluxes in excess of $\sim$\inflationflux\ \citep{demory11, Miller11}. Furthermore, a recent study by \citet{Sestovic18} suggests that above an incident flux of $\sim1.6 \times 10^{6}$ Wm$^{-2}$, all hot Jupiters in the mass range $0.37 - 0.98$ $M_J$ appear inflated. For statistical studies such as this, increasing the sample of well-characterised hot Jupiters is crucial, and allows us to further constrain and paramaterise inflation mechanisms, which remain poorly understood.

In this paper we report the discovery of four new hot Jupiter planets from the Next Generation Transit Survey (NGTS), three of which appear inflated. In \S2, we describe the discovery photometry from NGTS and the subsequent photometric and radial velocity follow-up observations. In \S3, we present our analysis of this data, including the determination of both the stellar and planetary parameters from spectral analysis, SED fitting, and global modelling. An investigation into the inflation of each planet is covered in \S4, with the subsequent results being discussed in \S5. Finally, our conclusions are laid out in \S6. \planetA, \planetB, \planetC\ and \planetD\ bring the total planet count from NGTS to 17 (note NGTS-7Ab is a brown dwarf).

\section{Observations}

\subsection{NGTS Discovery Photometry}
\label{sec:NGTSphot}

\begin{table*}
    \centering
    \caption{NGTS photometry for each object. The full tables are available in a machine-readable format from the online journal.  A portion is shown here for guidance.}
    \label{tab:ngtsphotsummary}
    \medskip
    \begin{subtable}{.40\linewidth}
        \centering
        \caption{NGTS photometry for \starA.}
        \smallskip
        \begin{tabular}{|c|c|c|}
            \hline
            Time	&	Flux        	&Flux\\
            (BJD-2450000)	&	(normalised)	&error\\
	        \hline
            7982.8536  &  0.973  &  0.031 \\
	        7982.8538  &  0.973  &  0.031 \\
	        7982.8539  &  1.001  &  0.031 \\
	        7982.8541  &  1.008  &  0.031 \\
	        7982.8542  &  1.023  &  0.031 \\
	        ...        &   ...    &   ...   \\
	        8195.5583  &  1.012  &  0.030 \\
	        8195.5584  &  1.039  &  0.030 \\
	        8195.5586  &  0.974  &  0.030 \\
	        8195.5587  &  0.959  &  0.030 \\
	        8195.5589  &  1.025  &  0.030 \\
	        \hline
        \end{tabular}
    \end{subtable}%
    \begin{subtable}{.40\linewidth}
        \caption{NGTS photometry for \starB.}
        \smallskip
        \centering
        \begin{tabular}{|c|c|c|}
            \hline
            Time	&	Flux        	&Flux\\
            (BJD-2450000)	&	(normalised)	&error\\
	        \hline
            7981.7857  &  1.010  &  0.031 \\
            7981.7860  &  0.968  &  0.031 \\
    	    7981.7862  &  0.985  &  0.031 \\
    	    7981.7863  &  1.007  &  0.031 \\
    	    7981.7865  &  1.008  &  0.031 \\
    	    ...        &   ...    &   ...   \\
    	    8191.5306  &  1.020  &  0.027 \\
    	    8191.5307  &  0.969  &  0.027 \\
    	    8191.5308  &  0.993  &  0.027 \\
    	    8191.5310  &  0.994  &  0.027 \\
    	    8191.5311  &  0.991  &  0.027 \\
    	    \hline
        \end{tabular}
    \end{subtable}
    
    \bigskip
    
    \medskip
    
    \begin{subtable}{.40\linewidth}
        \centering
        \caption{NGTS photometry for \starC.}
        \smallskip
        \begin{tabular}{|c|c|c|}
            \hline
            Time	&	Flux        	&Flux\\
            (BJD-2450000)	&	(normalised)	&error\\
    	    \hline
            7982.8537  &  1.001  &  0.025 \\
    	    7982.8539  &  1.021  &  0.025 \\
    	    7982.8540  &  1.014  &  0.025 \\
    	    7982.8542  &  1.002  &  0.025 \\
    	    7982.8543  &  1.023  &  0.025 \\
    	    ...        &   ...    &   ...   \\
    	    8195.5583  &  0.927  &  0.025 \\
    	    8195.5584  &  0.997  &  0.025 \\
    	    8195.5586  &  1.032  &  0.025 \\
    	    8195.5587  &  0.984  &  0.025 \\
    	    8195.5589  &  1.025  &  0.025 \\
    	    \hline
        \end{tabular}
    \end{subtable}%
    \begin{subtable}{.40\linewidth}
        \caption{NGTS photometry for \starD.}
        \smallskip
        \centering
        \begin{tabular}{|c|c|c|}
            \hline
            Time	&	Flux        	&Flux\\
            (BJD-2450000)	&	(normalised)	&error\\
    	    \hline
            8097.8232  &  1.016  &  0.032 \\
    	    8097.8233  &  0.961  &  0.032 \\
    	    8097.8235  &  0.977  &  0.032 \\
    	    8097.8236  &  1.007  &  0.032 \\
    	    8097.8238  &  1.005  &  0.032 \\
    	    ...        &   ...    &   ...   \\
            8334.5046  &  1.000  &  0.033 \\
        	8334.5047  &  0.991  &  0.033 \\
    	    8334.5049  &  1.010  &  0.033 \\
    	    8334.5050  &  1.026  &  0.033 \\
    	    8334.5052  &  0.983  &  0.033 \\
	        \hline
        \end{tabular}
    \end{subtable}% 
\end{table*}

\starA\ to \starD\ were initially identified as transiting exoplanet candidate systems from photometry from the Next Generation Transit Survey \citep[NGTS;][]{Wheatley18}, a ground-based wide-field exoplanet survey. Based at ESO's Paranal observatory in Chile, NGTS consists of an array of 12 independently mounted 20-cm Newtonian telescopes, each equipped with a 2K$\times$2K deep-depleted Andor IKon-L CCD camera. The custom NGTS 520-890\,nm filter optimises the survey for studies of K and M dwarf stars which, due to their small radii, provide the best opportunity for discovering small transiting planets.

\starA, \starB, and \starC\ were observed during the 2017 observing campaign for 160, 137, and 160 nights, respectively, between 2017 August 16 and 2018 March 18. \starD\ was observed the following season for 165 nights over the period of 2017 December 10 to 2018 August 7. Over 188,000 images were collected for each object using a single NGTS telescope with 10\,s exposures (see Table~\ref{tab:photsummary} for details).

The data were reduced and aperture photometry performed via the CASUTools package\footnote{\url{http://casu.ast.cam.ac.uk/surveys-projects/software-release}}, before being detrended using an optimised version of the SysRem algorithm \citep{Tamuz2005} which has been adapted for the NGTS pipeline. The data was then searched for transit-like events by \orion \citep{Wheatley18}, a custom implementation of the BLS fitting algorithm \citep{Kovacs2002}. \orion identified eight partial and three full transits for \starA; four partial and four full transits for \starB; 15 partial transits for \starC; and 16 partial and four full transits for \starD. Lightcurves of the NGTS detections, phase-folded on the best-fitting period, are shown in Figure~\ref{fig:allphotom}. The initial fits to the NGTS data provided by \orion revealed that the depths, widths and shapes of the transits for each object were compatible with transiting hot Jupiter planets. In addition, a convolutional neural network applied to the NGTS data found that the probabilities of each lightcurve containing a transiting exoplanet were all greater than 0.95, consistent with previous confirmed NGTS planet discoveries \citep{Chaushev19}.

In order to rule out the possibility that any of our detected companions were stellar rather than planetary, we performed additional checks on the NGTS data. The phase-folded lightcurves for each object were searched for any evidence of a secondary eclipse around phase 0.5 which would indicate the presence of a second star. Furthermore, we compared the transit depth of consecutive odd and even transits to check for a depth difference consistent with an eclipsing binary star system that had been mis-folded on half the true period. For all four targets, we find no indication in the NGTS photometry to suggest that the companions were not planets.

\subsection{Additional Photometry}

In order to confirm that each stellar companion was indeed a planet, and to constrain the transit parameters, we obtained additional photometry with the 1.0\,m and Lesedi telescopes at the South African Astronomical Observatory (SAAO). For three of the candidates, we were also able to use data from the Transiting Exoplanet Survey Satellite \citep{Ricker14}. The details of this additional photometry are outlined below.

\subsubsection{TESS}
\label{sec:TESSphot}

TESS is a space-based NASA survey telescope that searches for transiting planets around bright stars \citep{Ricker14}. It has a wide field of view, with four $24\times24^{\circ}$ cameras, each equipped with four 2k$\times$2k CCDs. Its typical observing baseline of 27 days makes it well-suited to detecting short-period transiting exoplanets.

We searched for our candidates in the TESS full frame images (FFIs) using the TESSCut\footnote{\url{https://mast.stsci.edu/tesscut/}} tool and found that \starB\ and \starC\ were observed in TESS Sectors 4 and 5, respectively, each for 27 consecutive nights between 2018 October 19 and 2018 December 11. In addition, \starD\ was observed between 2019 March 26 and 2019 April 22 in TESS Sector 10. \starA\ falls on a CCD in TESS Sector 5, but unfortunately into the overscan region of the camera rather than the science pixels, and therefore there is no TESS FFI observation for this star.  A summary of this information can be found in Table~\ref{tab:photsummary}.

Each star in the TESS FFIs was observed at 30 minute cadence, and for each of our candidates we used an automatically determined optimal aperture to exclude neighbouring stars. These aperture sizes ranged from three to eight pixels. However, the large pixel scale of TESS (21") meant that there was still some slight contamination of the FFI lightcurves, which we account for in our analysis (see Section \ref{sub:global}).

For each candidate, we found that the best BLS period from TESS was consistent with the \orion value for the NGTS photometry. TESS detected three full and two partial transits for \starB, eight full partial transits for \starC, and seven full transits for \starD. Phase-folded lightcurves of the TESS data can be found in Figure~\ref{fig:allphotom}.

\subsubsection{SAAO}
\label{sec:SAAOphot}

\begin{table*}
	\centering
	\caption{Summary of photometric observations.}
	\label{tab:photsummary}
	\begin{tabular}{ccccccc} 
    \hline
    \hline
Target & Instrument & Night(s) observed & N$_{\mathrm{images}}$ & Exptime (s) & Filter & Comments \\
\hline
\hline
        & NGTS  & 16/08/17 - 18/03/18	& 196732    & 10   & NGTS  & \\
\starA  & SAAO  & 21/11/18 	            & 285       & 60	    & I     & \\
        & SAAO	& 06/12/19	            & 276       & 30	    & I     & \\
\hline
\multirow{4}{*}{\starB} & NGTS	    & 16/08/17 - 18/03/18	& 188417	& 10   & NGTS  & \\
                        & TESS      & 19/10/18 - 14/11/18   & 831    & 1800*     & TESS & Sector 4\\
    	                & SAAO 1.0m	& 21/12/18	            & 200       & 60	    & I     & \\
                        & SAAO 1.0m	& 05/12/19	            & 540       & 30	    & I     & \\
\hline
        & NGTS	    & 16/08/17 - 18/03/18 	& 196732    & 10	& NGTS  & \\
\starC  & TESS	    & 15/11/18 - 11/12/18   & 1093    & 1800*	    & TESS & Sector 5\\
    	& SAAO 1.0m	& 05/02/19          	& 1104      & 15	    & I     & \\
\hline
\multirow{5}{*}{\starD}	& NGTS	    & 10/12/17 - 07/08/18 	& 243515    & 10	& NGTS  & \\
                        & SAAO 1.0m	& 21/03/19 	            & 598       & 20	    & I     & \\
                        & TESS	    & 26/03/19 - 22/04/19   & 1017    & 1800*	    & TESS & Sector 10\\
                        & Lesedi	& 24/06/2020            & 426       & 15        & I     & \\
                        & Lesedi	& 30/06/2020            & 550       & 15	    & V     & \\
	\hline
    \multicolumn{7}{l}{The NGTS filter has a bandpass from 520-890 nm; the TESS bandpass spans 600-1000nm.} \\
    \multicolumn{6}{l}{*TESS cameras have an exposure time of 2s but are stacked to 30min cadence.}\\
    \end{tabular}
\end{table*}

Between 2018 November and 2019 December, we took at least one set of follow-up photometry for each target using the 1.0\,m telescope at SAAO. The telescope was equipped with the full-frame transfer CCD Sutherland High-speed Optical Camera (SHOC), ``\textit{SHOC'n'awe}", and all observations were taken using the \textit{I} band filter. The $2.85' \times 2.85'$ field of view of \textit{SHOC'n'awe} on the 1.0\,m telescope allowed us to simultaneously observe at least one comparison star of similar brightness for each object.

For \starD, we also had the opportunity to obtain two lightcurves using the newly commissioned 1.0\,m Lesedi telescope at SAAO, which was equipped with the ``\textit{SHOC'n'disbelief}'' optical camera. The instrument's slightly wider field of view ($5.7' \times 5.7'$) allowed for an increased selection of comparison stars.

\medskip

Two lightcurves of \starA\ were obtained using the 1.0\,m telescope on the nights of 2018 November 21 and 2019 December 6 with exposure times of 60s and 30s, respectively. We obtained 285 images for the 2018 data, and truncated the 2019 data from 543 to 276 frames due to extremely poor conditions towards the end of the night. Similarly, additional photometry was also obtained using the same telescope on the night of 2020 February 3, but a combination of poor atmospheric conditions (with seeing reaching $6.5''$) and load shedding (scheduled electrical power shutdowns) at SAAO caused data gaps and fluctuations in the lightcurve on the order of the size of the transit signal, and so we omit this data from our analysis.

\begin{figure*}
    \centering
    \begin{subfigure}[b]{0.48\linewidth}
    \centering
	\includegraphics[width=0.98\columnwidth]{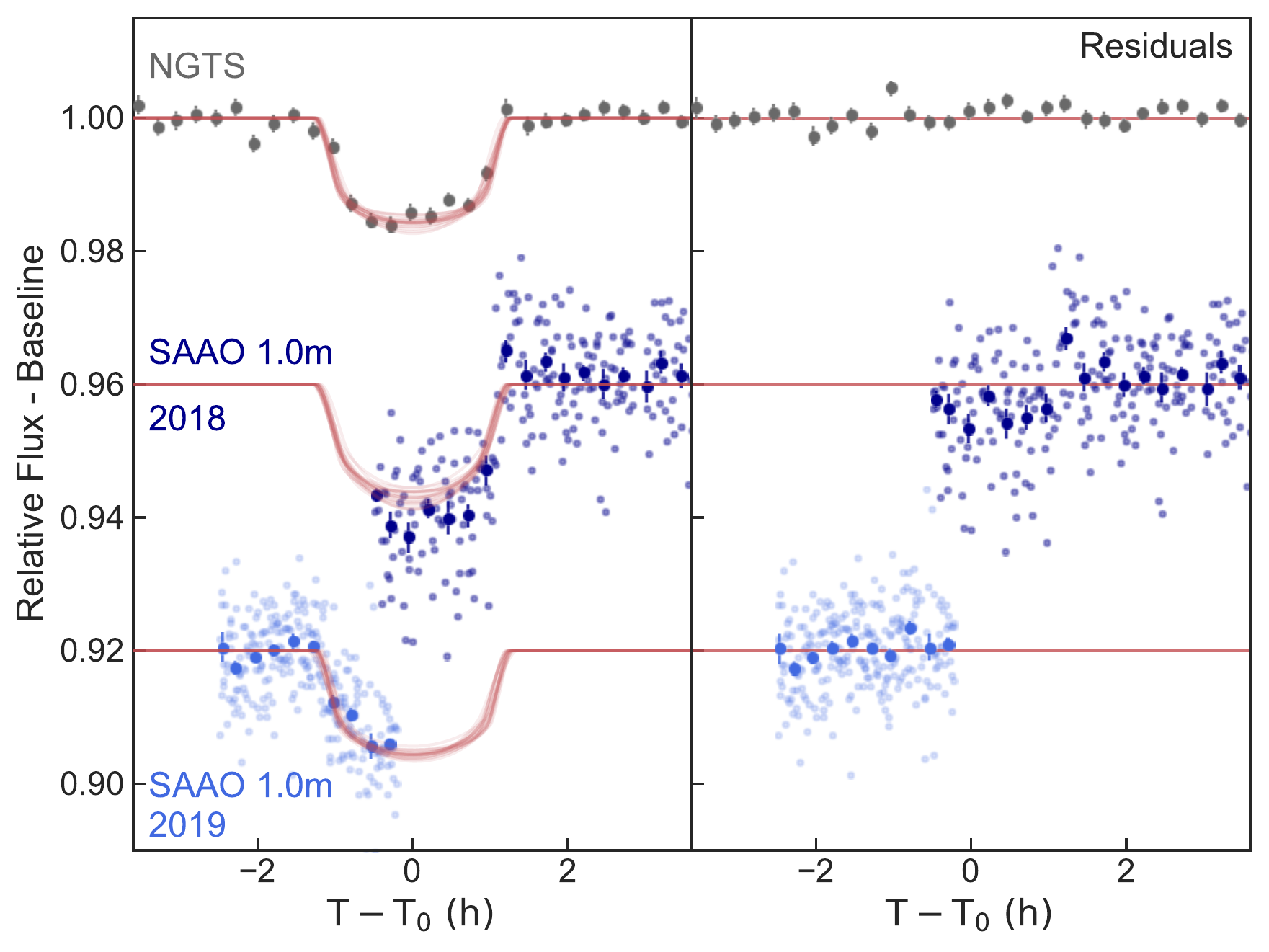}
	\caption{\textbf{Photometry for \planetA.} The \NGTS\ discovery lightcurve is phase-folded at the best-fitting period of \Abperiod\,d. The lightcurves from the 1.0\,m telescope at SAAO show a detection of egress from the 21st November 2018 and a detection of ingress from 6th of December 2019. Note that the  in-transit scatter during the 2018 observation is a result of poor atmospheric conditions, and that the transit depths from global modelling for the NGTS and SAAO lightcurves agree to within errors. \\}
    \label{fig:planetA_photom}
    \end{subfigure}%
    \hspace{0.15in}
    \begin{subfigure}[b]{0.48\linewidth}
    \centering
	\includegraphics[width=0.98\columnwidth]{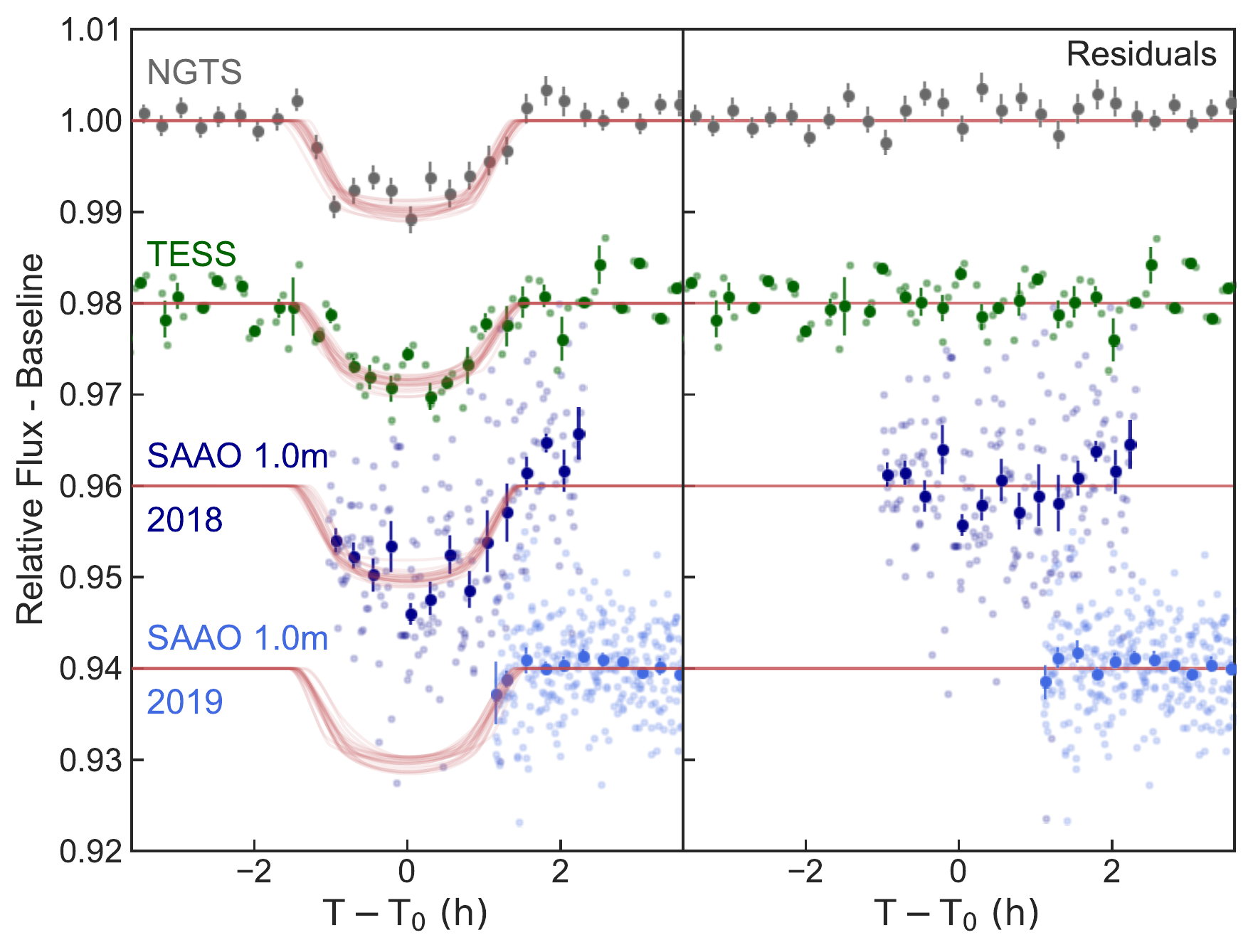}
	\caption{\textbf{Photometry for \planetB.} The \NGTS\ and TESS lightcurves are phase-folded on the best fitting period, \Bbperiod\,d. For this object, we obtained two lightcurves with the SAAO 1.0\,m telescope on 21st December 2018 and 5th December 2019, both of which include egress. As with \starA, the large in-transit scatter of the SAAO lightcurve from 2018 is due to poor atmospheric conditions and the true errors of these data points are likely underestimated. All four transit depths from the global modelling agree to within errors.}
    \label{fig:planetB_photom}
    \end{subfigure}
    
    \bigskip
    
    \begin{subfigure}[b]{0.48\linewidth}
    \centering
	\includegraphics[width=0.98\columnwidth]{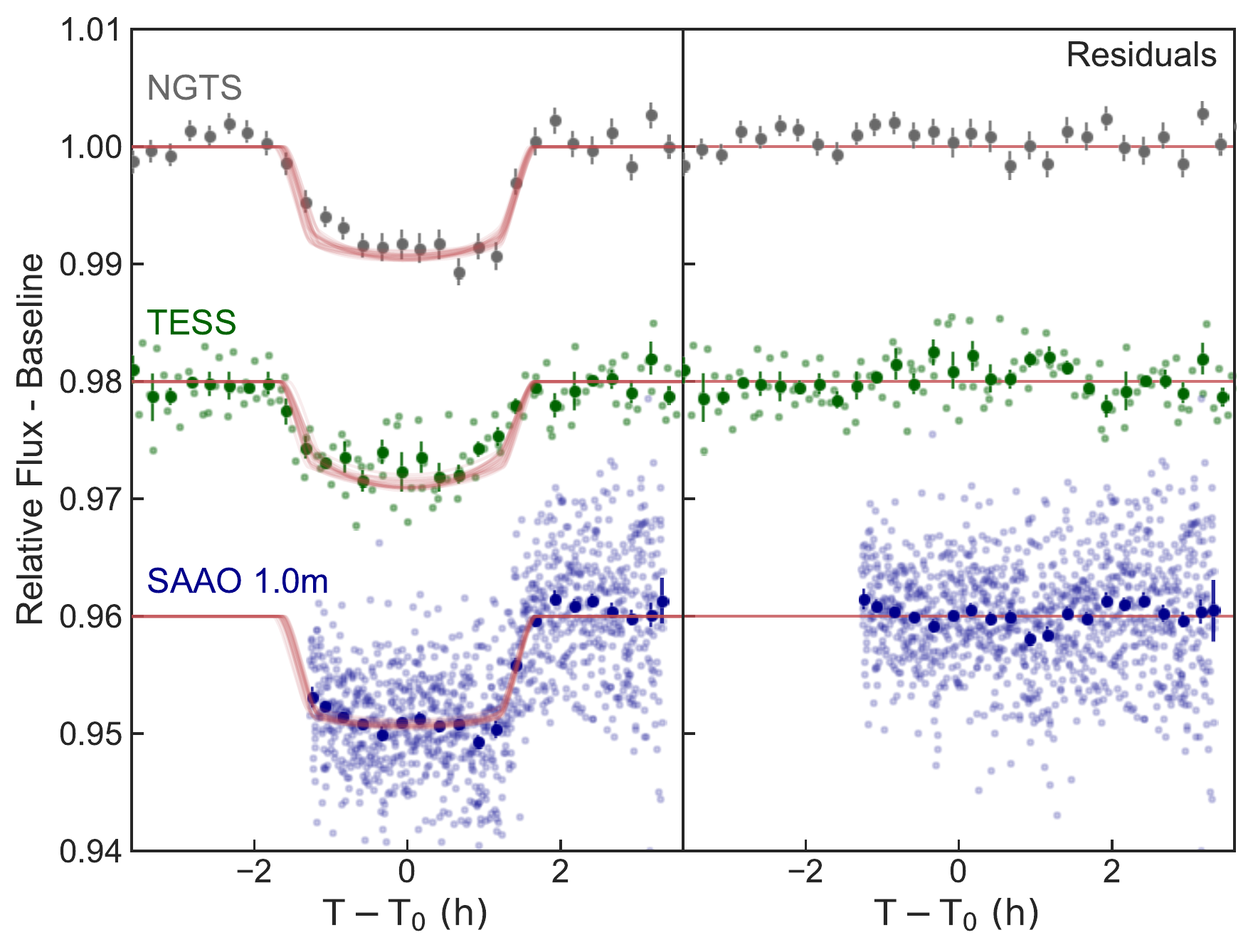}
	\caption{\textbf{Photometry for \planetC.} The \NGTS\ discovery lightcurve and TESS follow-up lightcurve are phase-folded at the best-fitting period of \Cbperiod\,d. The SAAO lightcurve consists of one dataset from 5th February 2019.\vspace{0.4148in}}
    \label{fig:planetC_photom}
    \end{subfigure}%
    \hspace{0.15in}
    \begin{subfigure}[b]{0.48\linewidth}
    \centering
	\includegraphics[width=0.98\columnwidth]{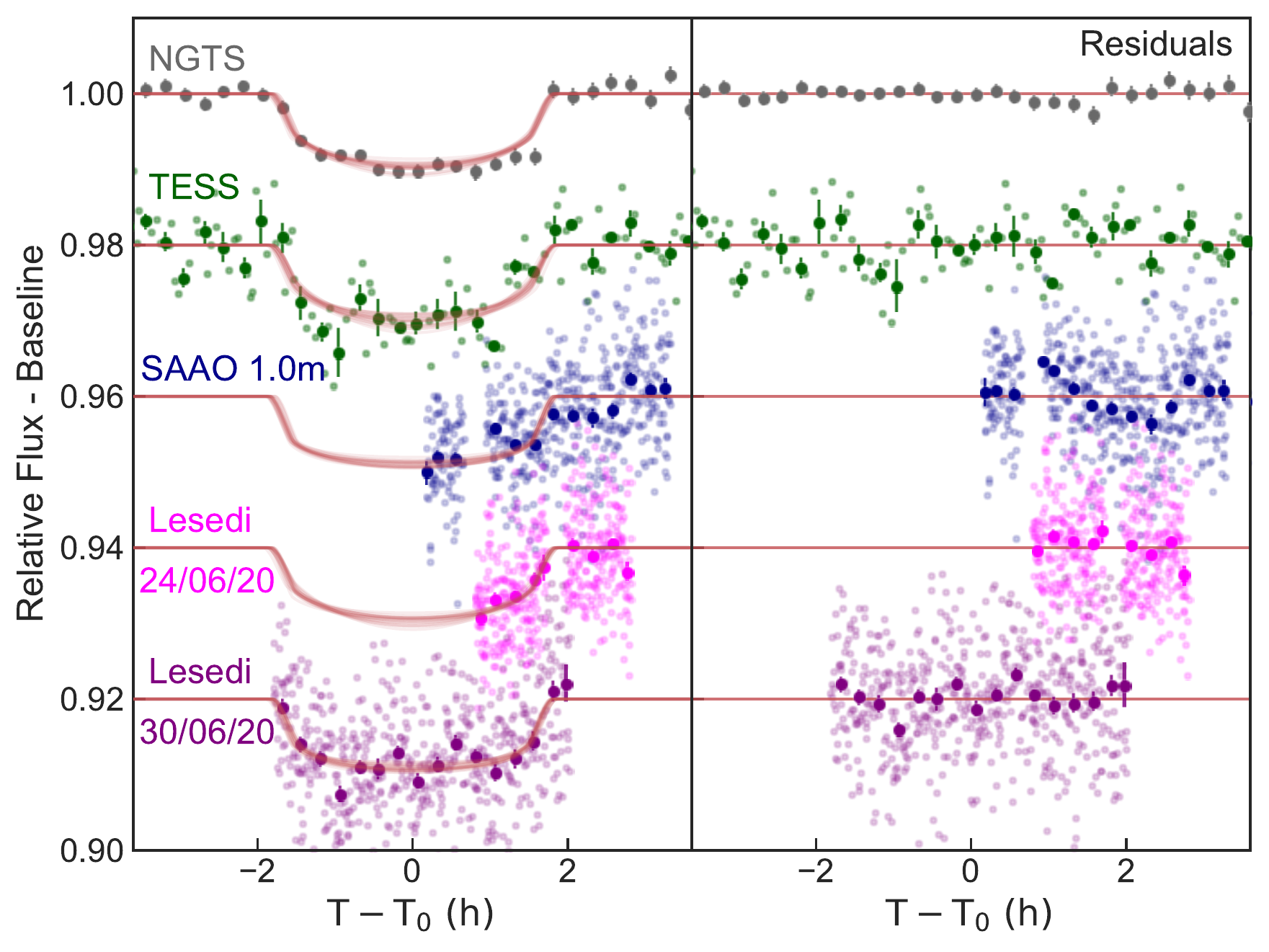}
	\caption{\textbf{Photometry for \planetD.} The \NGTS\ discovery lightcurve and TESS follow-up lightcurve are phase-folded at the best-fitting period of \Dbperiod\,d. The SAAO 1.0\,m telescope lightcurve consists of one dataset from 21st March 2019; although egress is seen, due to the poor quality of the data, we omit it from our global modelling. The two Lesedi datasets in \textit{I} and \textit{V} band comprise of single observations from 24th and 30th June respectively.}
    \label{fig:planetD_photom}
    \end{subfigure}
    \caption{NGTS discovery photometry (grey points) and follow up photometry (TESS: green points, SAAO 1.0\,m telescope \textit{I} band: dark and light blue points, Lesedi \textit{I} band: pink points, Lesedi \textit{V} band: purple points) for each planet. Note that, for clarity, we exclude the unbinned NGTS data from the plots. The red lines show 20 lightcurve models generated from randomly drawn posterior samples of the \allesfitter fit. Residuals are shown to the right of the lightcurves.}
    \label{fig:allphotom}
\end{figure*}

Load shedding also tainted the observations for \starD\ in 2019 March, resulting in a noticeable in-transit data gap of about 20 minutes in the final lightcurve. Despite being able to identify transit egress in this data, we omitted it from our final modelling process, as the data quality was too poor to obtain a reliable fit. However, we include the lightcurve in Figure~\ref{fig:planetD_photom}. We instead utilise follow up photometry from the Lesedi telescope, taken on the nights of 2020 June 24 and 30 in the \textit{I} and \textit{V} bands, respectively. The \textit{I} band data consists of 426 $\times$ 15\,s exposures; although it also includes a data gap of about 14\,min long, this time due to an auto guider failure, the in-transit data is stable, and we thus include it in the modelling. Fortunately, the final dataset in \textit{V} band was taken continuously over 2.29\,hr, and consists of 550 $\times$ 15s exposures. 

\starB\ and \starC\ were both observed using only the 1.0\,m telescope at SAAO.\@ For \starB, we obtained two follow-up lightcurves on the night of 2018 December 21 (200 $\times$ 60s exposures) and the night of 2019 December 5 (540 $\times$ 30s exposures), whilst \starC\ was observed once on the night of 2019 Febuary 5 (1104 $\times$ 15\,s exposures).

A full summary of the photometric observations for each object is detailed in Table~\ref{tab:photsummary}.

\medskip

Each lightcurve was bias and flat-field corrected using the local \textsc{python}-based SAAO SHOC pipeline, which uses \textsc{iraf} photometry tasks (\textsc{pyraf}) and facilitates the extraction of raw and differential lightcurves. We used the Starlink package \textsc{autophotom} to perform aperture photometry on both our target and comparison stars, and chose apertures that gave the maximum signal-to-noise ratio. Background apertures were adjusted to account for changes in apparent star size over the night as the atmospheric conditions varied. Finally, the measured fluxes of the comparison stars for each object were used for differential photometry of our targets.

\medskip

The SAAO lightcurves of each candidate are shown in Figure~\ref{fig:allphotom}. For \starA\ and \starB, which each have two SAAO detections from the 1.0\,m telescope, the data has been phase-folded on the best-fitting period.

We detected a clear egress for \starA\ in 2018 November, as well as an ingress in 2019 December. Although seeing reached 6.4" near the beginning of the 2018 observations, resulting in large in-transit scatter for this lightcurve, our fitting procedure reveals that the transit depth is consistent with the NGTS data to within errors. Additionally, we note that these errors are likely underestimated for the weather-affected parts of the SAAO lightcurve.

Analysis of the 2018 December data for \starB\ revealed the majority of a transit, just missing ingress, whilst the 2019 December data is primarily out-of-transit with a few data points in egress. As with the 2018 data for \starA, the 2018 data for \starB\ was affected by varying atmospheric conditions, with the full moon and high cloud causing in-transit scatter. However, again the transit depths between telescopes are in agreement.

Similar the 2018 lightcurve for \starB, we detected close to a full transit for \starC, again just missing ingress. Finally, the data for \starD\ from the Lesedi telescope contained an egress on the night of 2020 June 24 and a full transit, including some points in egress, from the night of 2020 June 30.

For all four objects, the transits observed with SAAO were consistent with planetary companions. Subsequently each were flagged for spectroscopic follow-up to enable mass determination.

\subsection{Spectroscopic Follow-Up}
\label{sec: specFU}

\begin{table*}
	\centering
	\caption{Summary of radial velocity measurements. The full tables can be found in Appendix \ref{sec:appendix}.}
	\label{tab:rvsummary}
	\begin{tabular}{ccccccc} 
    \hline
    \hline
    Target & Instrument & Nights observed & N$_{\mathrm{spectra}}$ & Exptime (s) & Programme & SNR$_{\mathrm{combined}}$ \\
\hline
\hline
        & CORALIE   & 01/11/2018 - 01/03/2019   & 7     & 2700  &    N/A & - \\
\starA	& HARPS     & 13/09/2019 - 03/12/2019	& 7     & 1800  &   0103.C-0719(A) \& 0104.C-0588(A) & 16.80 \\
        & FEROS	    & 10/09/2019 - 19/09/2019	& 7     & 1800  &   0103.A-9004(A) & - \\
\hline
\multirow{2}{*}{\starB}	& HARPS     & 21/01/2020 - 22/03/2020	& 6     & 1800	&   0104.C-0588(A) & 21.77 \\
        & FEROS	    & 30/12/2019 - 04/01/2020	& 6     & 1800	&   0104.A-9012(A) & - \\
\hline
\multirow{2}{*}{\starC}  & CORALIE   & 01/10/2019 - 14/02/2020   & 8     & 2700  &  N/A  & - \\
        & FEROS	    & 09/11/2019 - 04/01/2020	& 12    & 1800  &   0103.A-9004(A) \& 0104.A-9012(A) & 59.00 \\
\hline
        & CORALIE   & 21/05/2019 - 14/01/2020   & 2     & 2700          &  N/A & - \\
\starD	& HARPS     & 02/03/2020 - 23/03/2020	& 10    & 2400 - 2700   &   0104.C-0588(A) & 26.69 \\
        & FEROS	    & 30/12/2019 - 04/01/2020	& 5     & 1800          &   0104.A-9012(A) & - \\
	\hline
    \end{tabular}
\end{table*}

\begin{figure*}
    \begin{subfigure}[b]{0.45\linewidth}
    \centering
	\includegraphics[width=0.95\columnwidth]{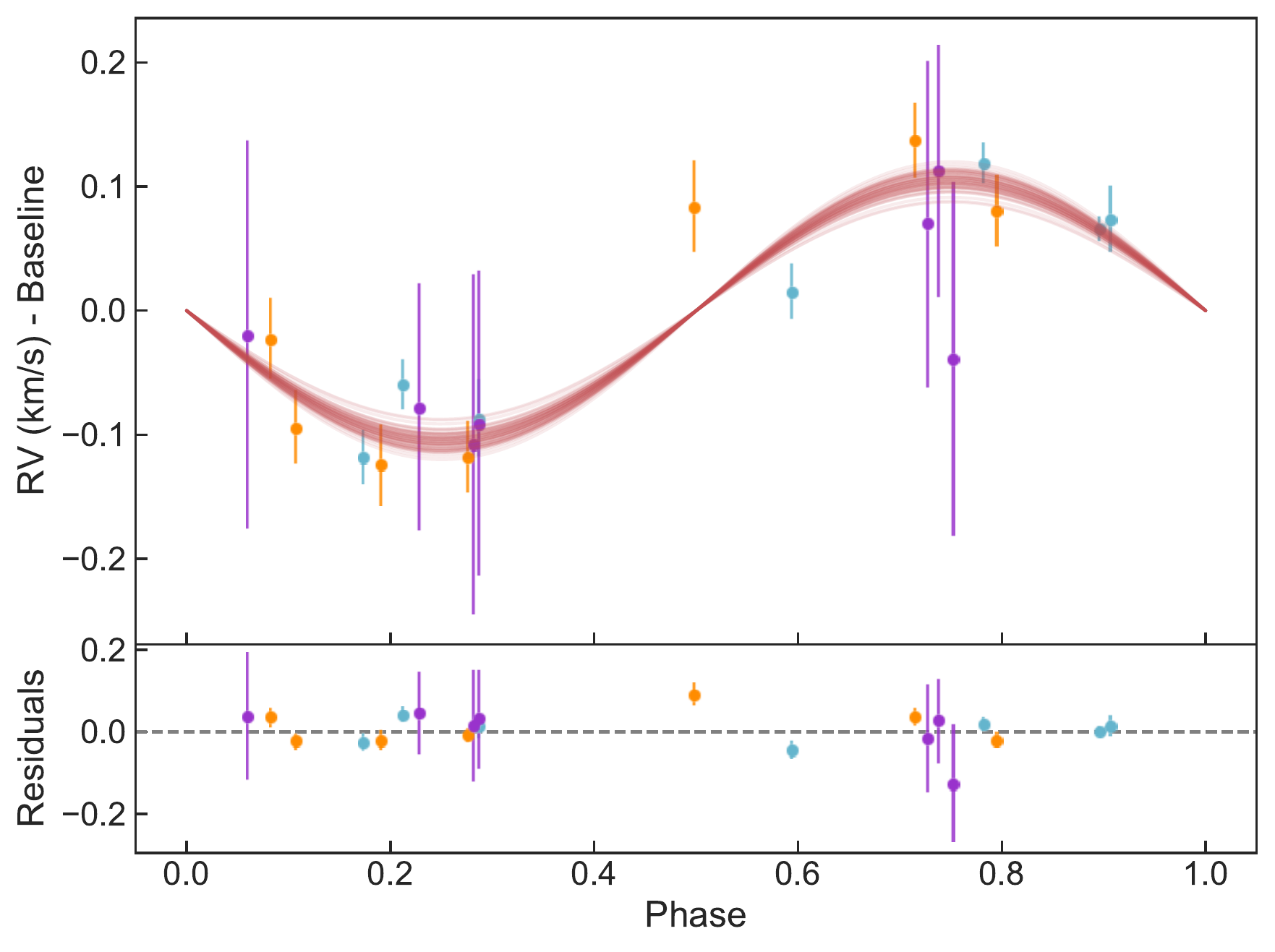}
	\caption{\starA}
    \label{fig:starARVs}
    \end{subfigure}%
    \hspace{0.15in}
    \begin{subfigure}[b]{0.45\linewidth}
    \centering
	\includegraphics[width=0.96\columnwidth]{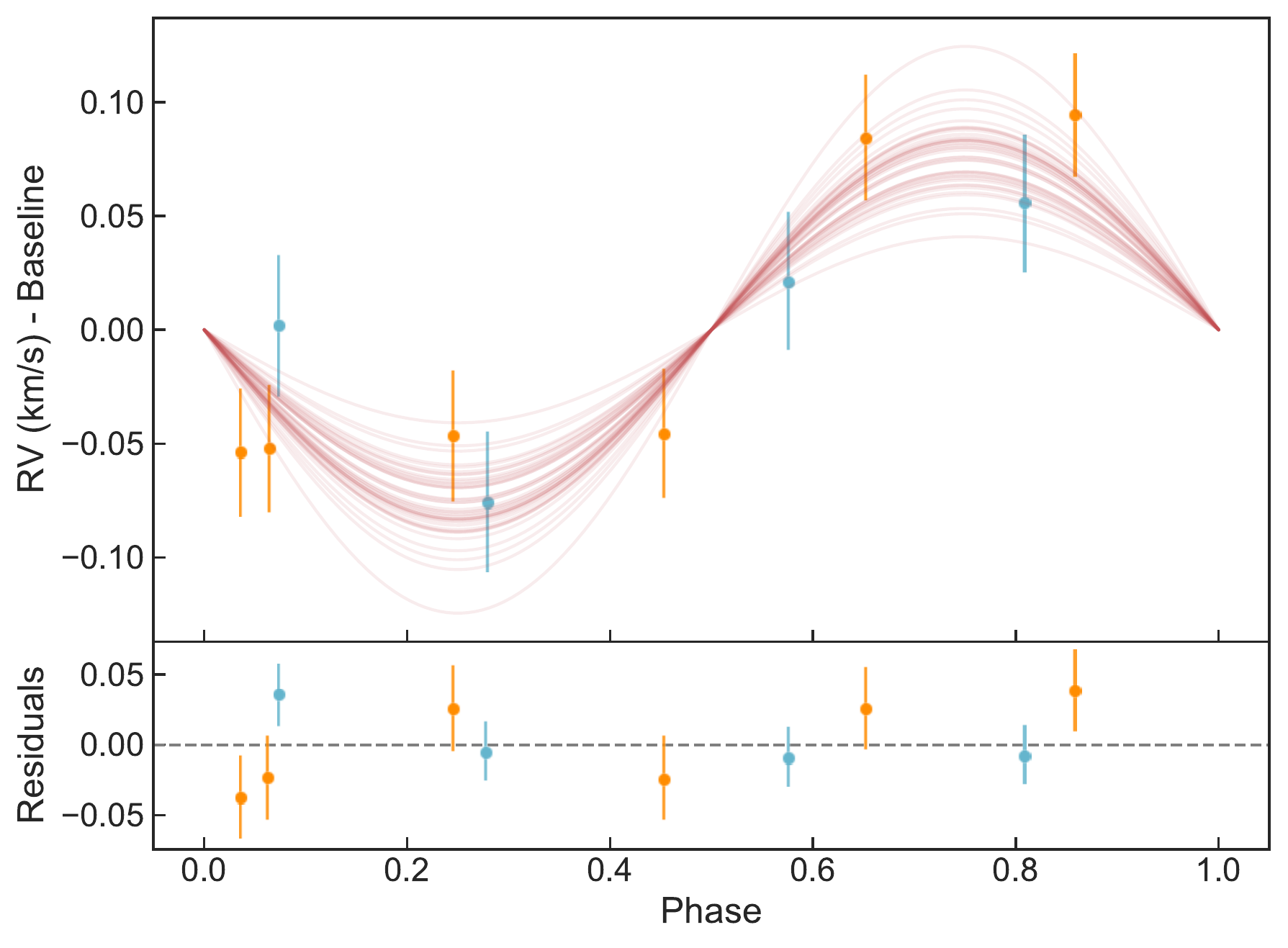}
	\caption{\starB}
    \label{fig:starBRVs}
    \end{subfigure}
    
    \bigskip
    
    \begin{subfigure}[b]{0.45\linewidth}
    \centering
	\includegraphics[width=0.95\columnwidth]{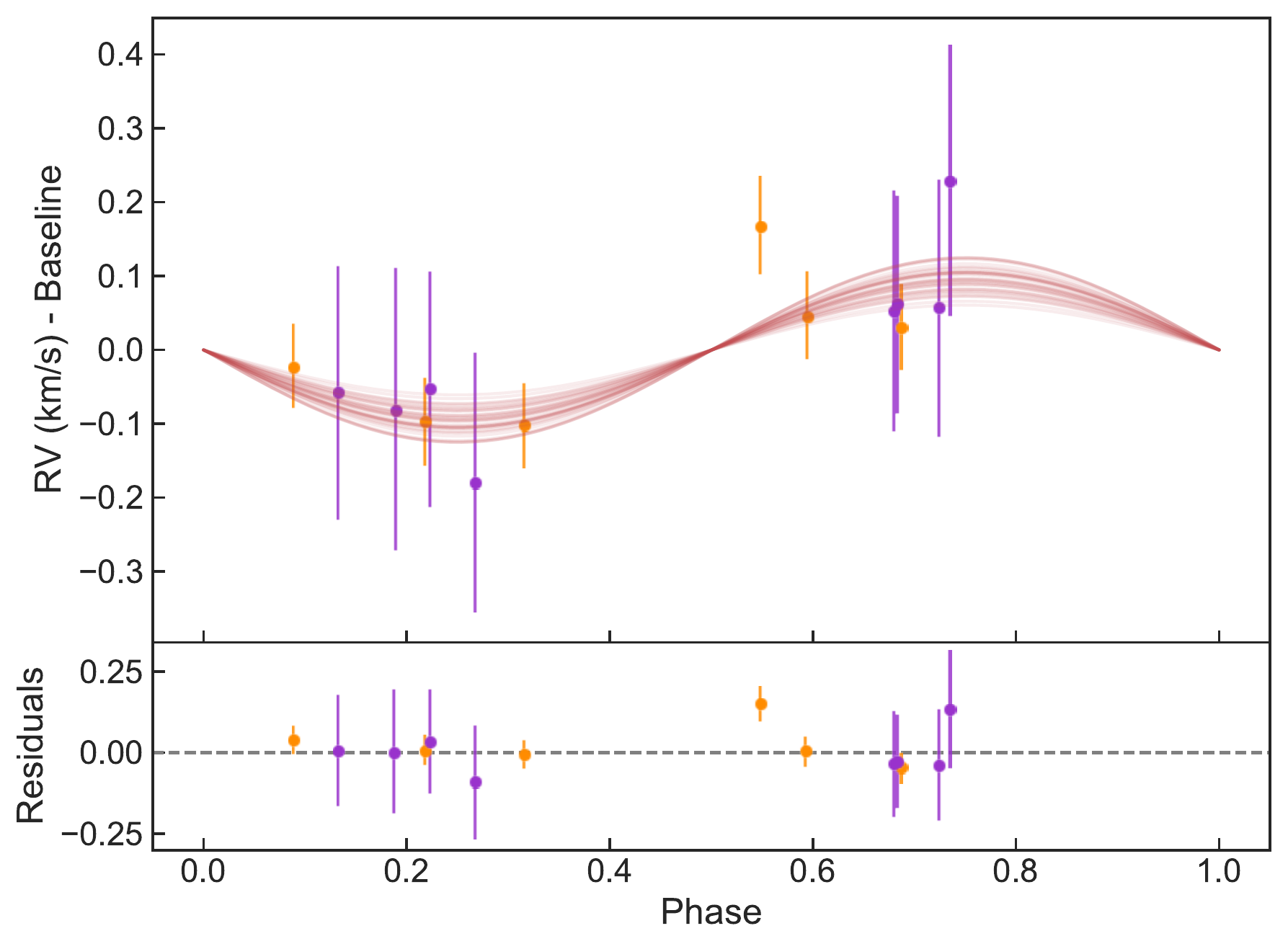}
	\caption{\starC}
    \label{fig:starCRVs}
    \end{subfigure}%
    \hspace{0.15in}
    \begin{subfigure}[b]{0.45\linewidth}
    \centering
	\includegraphics[width=0.95\columnwidth]{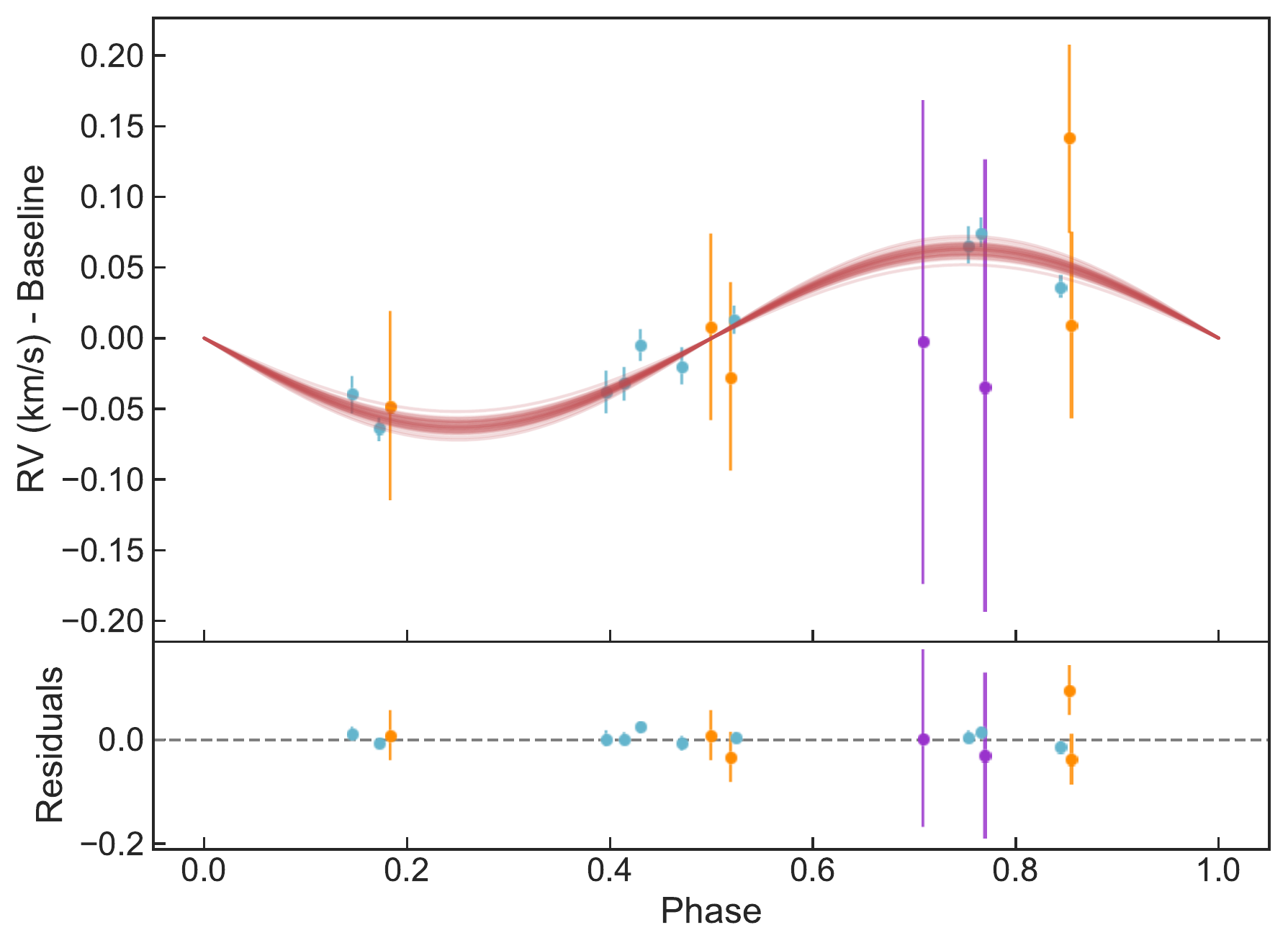}
	\caption{\starD}
    \label{fig:starDRVs}
    \end{subfigure}
    \caption{Phase-wrapped radial velocity measurements for all four stars from the HARPS (cyan points), FEROS (orange points) and CORALIE (purple points) spectrographs. The red lines show 20 radial velocity models generated from randomly drawn posterior samples of the \allesfitter fit.}
    \centering
    \label{fig:allRVs}
\end{figure*}

In order to constrain the masses of the planetary companions, we obtained multi-epoch spectroscopy for radial velocity measurements. Three different fibre-fed \'echelle spectrographs were used, all located at the La Silla Observatory in Chile. The HARPS spectrograph \citep{mayor03} is mounted on the ESO 3.6\,m telescope, CORALIE \citep{Queloz01CORALIE} on the Swiss 1.2m Leonard Euler telescope, and FEROS \citep{Kaufer98} on the 2.2m MPG/ESO telescope. The spectrographs have a range of spectral resolutions of R = 115,000, R = 60,000 and R = 48,000, respectively. Due to the faintness of the targets (14.3 < V < 14.6; see Tables \ref{tab:stellar103524}, \ref{tab:stellar103406}, \ref{tab:stellar103867}, and \ref{tab:stellar104605}), all HARPS spectra were collected with the high efficiency fibre link (EGGS), which uses a fibre size of 1.4\,\arcsec\ instead of the usual 1.0\,\arcsec\ mode. The HARPS-EGGS mode trades spectral resolution for approximately twice the photon count, depending on seeing. We used the second HARPS-EGGS fibre to monitor the sky simultaneously with science observations. 
All four objects were observed by FEROS and at least one other spectrograph, and we obtained a minimum of 12 data points for each (see Table~\ref{tab:rvsummary} for details).

The HARPS data were reduced via the offline data reduction pipeline (DRS) before being cross-correlated with a binary G2 mask to extract radial velocities \citep{baranne96}. This procedure was also employed for the CORALIE data, for which the spectra were reduced using the standard CORALIE DRS. For the data collected with FEROS, the CERES reduction pipeline \citep{Brahm17} performed a radial velocity extraction by the same method.

Two HARPS spectra of \starD\ obtained on BJD 2458916.780 and 2458918.843 were contaminated by moon light. We corrected the radial velocity measurements by subtracting the signal of the simultaneous sky-fibre from the science fibre in cross correlation function (CCF) space. The RVs are then extracted as per usual on the star-sky CCF. Additionally, the CORALIE radial velocities for \starC\ were computed while excluding the first 30 spectral orders, in which the signal-to-noise ratio was $<1$. Phase-folded plots of the radial velocity data for all four targets are shown in Figure \ref{fig:allRVs}.

For all four targets, the radial velocity measurements were consistent with a Jupiter-mass planet, with semi-amplitudes on the order of K$\sim$100\,\ms (see Table \ref{tab:planetparams}) and variations in-phase with the periods derived from the photometric data. To confirm that the signals did not arise due to cool stellar spots or a blended eclipsing binary, we checked for any correlation between the radial velocity measurements and the line bisector span of the CCF \citep{queloz01CCF}. Correspondingly, no evidence was found in the spectroscopic data that contradicted the existence of a planetary companion.

\section{Analysis}

\subsection{Stellar Properties}
\label{sub:stellar}

In order to produce reliable fits to our photometric and radial velocity data for \starA, \starB, \starC\ and \starD, we must first determine the stellar properties of each system. Constraining these is vital as the accuracy of the measured planetary parameters is dependant on how well we can characterise the host star. Below we outline our stellar analysis procedure and results.

\subsubsection{Gaia DR2}
\label{subsub:Gaia}

For all four objects, we obtained additional photometric and astrometric data from \textit{Gaia DR2} \citep{GAIA, Gaia2018b} (see Tables \ref{tab:stellar103524}, \ref{tab:stellar103406}, \ref{tab:stellar103867}, and \ref{tab:stellar104605}). Part of these data, including the stellar radius and parallax, were utilised as priors in our SED fitting procedure (see Section \ref{sub:ariadne}).

The \textit{Gaia} parallaxes, proper motions, and absolute radial velocities of each star were used to determine the Galactic velocity components (\textit{U$_{LSR}$,V$_{LSR}$,W$_{LSR}$}), assuming a Local Standard of Rest of UVW$=$(11.1, 12.14, 7.25)km s$^{-1}$ \citep{Schonrich2010}. By using the selection criteria for kinematically thin-disk ($V_{\mathrm{tot}}<50$ kms$^{-2}$) and kinematically thick-disk ($70<V_{\mathrm{tot}}<180$ kms$^{-2}$) objects \citep{Gaia2018b}, we conclude that all four stars belong to the thin disk population (see Table \ref{tab:kinematics}).

\begin{table*}
	\centering
	\caption{The galactic velocity components and total space velocities for NGTS-15 - 18.}
	\label{tab:kinematics}
	\begin{tabular}{ccc}
    \hline
    Target & (\textit{U$_{LSR}$,V$_{LSR}$,W$_{LSR}$}) & $V_{\mathrm{tot}}$ (\kms) \\
\hline
    NGTS-15 & (\Agalu, \Agalv, \Agalw) & 32 \\
    NGTS-16 & (\Bgalu, \Bgalv, \Bgalw) & 28 \\
    NGTS-17 & (\Cgalu, \Cgalv, \Cgalw) & 32 \\
    NGTS-18 & (\Dgalu, \Dgalv, \Dgalw) & 16 \\
    \hline
    \end{tabular}
\end{table*}

Finally, our host stars show no evidence of being unresolved binaries, with all four targets having an astrometric excess noise of zero as well as a low ($<0$) astrometric goodness of fit in the along-scan direction (\texttt{GOF\_AL}) in \textit{Gaia DR2} \citep{Evans18}.

\subsubsection{Stellar analysis}

Although we used spectral energy distribution (SED) fitting to derive our final stellar parameters (see Section~\ref{sub:ariadne}), we first fit the stacked spectra for each object in order to obtain suitable priors for the key parameters with which to constrain the SED fits. For this we used \ispec \citep{Blanco14, blanco19}, an open-source framework for spectral analysis. \ispec users can choose from a list of models for spectral synthesis; we employ the \textsc{spectrum} \citep{gray94} radiative transfer code and the solar abundance model from \cite{asplund09}. Additionally, we use the Gaia-ESO Survey (GES) line list (version 5.0 -- \citealt{heiter15}), which covers the full wavelength range of our spectra (from 420-920nm), and the MARCES.GES model atmosphere \citep{Gustafsson08}.

We used our high-resolution HARPS data for the spectral analysis of \starA, \starB, and \starD. Since \starC\ was not observed with HARPS, we instead used the FEROS spectra.  For each object, we shift the spectra to the laboratory frame of reference and co-add them to produce a single, high SNR, combined spectrum. These SNRs are reported in Table \ref{tab:rvsummary}. We note that, although \starC\ has the highest SNR, the data was affected by correlated red noise, which hampered spectral analysis.

We determined the stellar effective temperature, \teff, and the surface gravity, \logg, via fits to the the H$\alpha$, NaI D and MgI b lines. Individual FeI and FeII lines were used as diagnostics for metallicity (\feh) and the rotational line-of-sight broadening (\vsini). Values for each parameter were obtained by synthesising model spectra until we found an acceptable fit to the data, and uncertainties were estimated by varying the fit until the models no longer matched the stacked spectra of each object.

From our spectral analysis, we find all four stars to be metal-rich, which aligns with observations that short and intermediate period gas giants tend to be found around more metal-rich stars than longer period gas giants \citep{jenkins17, Maldonado2019}.

\subsubsection{SED fitting}
\label{sub:ariadne}

To determine the stellar parameters of each system, we performed a fit to their SEDs.  For this, we employed the python tool \ariadne \citep{Vines21}, which fits catalogue photometry to different atmospheric model grids (\texttt{Phoenix V2} \citep{Husser13}, \texttt{BT-Settl}, \texttt{BT-Cond}, \texttt{BT-NextGen} \citep{Allard12, Hauschildt99}, \citet{Castelli04}, \citet{Kurucz93}) which are then convolved with a range of filter response functions.

The SEDs were modelled by interpolating the model grids in \teff\--\logg\--\feh\ space, with radius, distance, and extinction in the $V$ band used as additional model parameters. Any underestimated uncertainties in the photometry were accounted for by including an individual excess noise term for each photometric dataset.

Priors for \teff, \logg, and \feh\ were taken from our stellar analysis with \ispec, whilst priors for the stellar radius and distance were obtained from the \textit{Gaia DR2} values, although we used an inflated value for the radius error to account for modelling errors. In addition, $A_V$ was limited to the maximum line-of-sight value taken from the re-calibrated SFD galactic dust map \citep{Schlegel98, Schlafly11}. The priors for the excess noise parameters were set to a normal distribution with a mean of zero and a variance equal to five times the associated uncertainty.

The bayesian evidence for each model was calculated using the \dynesty nested sampling package, which also produces the parameters' posterior samples  \citep{speagle20}. The relative probabilities of the models were then used as weights with which to calculate the weighted average of each parameter via the following equation:

\begin{equation}
    P(\theta_i) = \sum_{n=1}^N P(\theta | X, M_n)P(M_n |X)
\end{equation}

\noindent Where $\theta_i$ is the parameter to be averaged, $P(\theta | X,M_n)$ is the posterior distribution derived using Bayes Theorem, and $P(M_n | X)$ is the Bayesian evidence of the individual model.

Note that, by averaging over several posterior distributions, \ariadne is able to achieve a higher precision than would typically be obtained with a single atmospheric model. Subsequently, some uncertainties may be underestimated. We therefore calculated an additional systematic error for these parameters by following the approach taken in \citet{Southworth2015}, wherein each SED was fit with the individual stellar atmosphere models before being compared to the overall Bayesian model averaging solution. By taking the largest difference between the model values and the averaged value from all posterior distributions, we were able to obtain the systematic uncertainty.

Finally, the mass value was estimated using MIST isochrones \citep{Choi16}. The derived stellar parameters are listed in Tables \ref{tab:stellar103524}, \ref{tab:stellar103406}, \ref{tab:stellar103867} and \ref{tab:stellar104605}. A detailed overview of the \ariadne fitting procedure, as well as the accuracy and precision of the tool, can be found in \citet{Vines21}.

\begin{table}
    \renewcommand{\arraystretch}{1.25}
	\centering
	\caption{Stellar Properties for \starA. For the values from \ariadne, we include a second systematic error from the difference between the average best-fit value and the maximum value from the individual stellar atmosphere models.}
	\begin{tabular}{lcc} 
	\hline
	Property	&	Value		&Source\\
	\hline
	2MASS I.D.	& 04532526-3248011	&2MASS	\\
    Gaia source I.D. & 4873830691665395584 & \textit{Gaia DR2} \\
    TIC I.D. & TIC-1333933 & TIC8\\
    \\
    \multicolumn{3}{l}{Astrometric Properties}\\
    R.A. &	\ARA	&\textit{Gaia DR2}	\\
	Dec	& \ADec	&\textit{Gaia DR2}	\\
    $\mu_{{\rm R.A.}}$ (\masy) & \ApropRA & \textit{Gaia DR2} \\
	$\mu_{{\rm Dec.}}$ (\masy) & \ApropDec & \textit{Gaia DR2} \\
	Parallax (mas)             & \Aparallax & \textit{Gaia DR2} \\
    \\
    \multicolumn{3}{l}{Photometric Properties}\\
	V (mag)		&\AVmag 	&APASS\\
	B (mag)		&\ABmag		&APASS\\
	g (mag)		&\Agmag		&APASS\\
	r (mag)		&\Armag		&APASS\\
	i (mag)		&\Aimag		&APASS\\
    G (mag)		&\AGAIAmag	& \textit{Gaia DR2}\\
    J (mag)		& \AJmag &2MASS	\\
   	H (mag)		& \AHmag &2MASS	\\
	K (mag)		& \AKmag	&2MASS	\\
    W1 (mag)	&\AWmag		&WISE	\\
    W2 (mag)	&\AWWmag	&WISE	\\
    T (mag)     &\ATESSmag  &TIC8   \\
    \\
    \multicolumn{3}{l}{Derived Properties}\\

    Spectral type                           & \Aspectype    & \ariadne \\
    T$_{\rm eff}$ (K)                       & \Ateff        & \ispec\\
    \feh             		& \Ametal       & \ispec\\
    vsini (\kms)	                        & \Avsini	    & \ispec\\
    $\gamma_{RV}$ (\kms)    & \Agamma		& RV data \\
    log g                                   & \Alogg	    & \ariadne\\
    \mstar (\msun)                          & \Astarmass    & \ariadne\\
    \rstar (\rsun)                          & \Astarradius  & \ariadne\\
    Age (Gyrs)		  & \Aage				        	&SED fitting\\
    Distance (pc)	        &  \Adist	    & \ariadne \\
    $A_v$ (mag)                             & \AAv     & \ariadne \\
    
	\hline
    \multicolumn{3}{l}{2MASS \citep{2MASS}; APASS \citep{APASS};}\\
    \multicolumn{3}{l}{WISE \citep{WISE}; {\em Gaia} \citep{GAIA};}\\
    TIC8 \citep{Stassun19}\\
	\end{tabular}
    \label{tab:stellar103524}
\end{table}

\begin{table}
    \renewcommand{\arraystretch}{1.25}
	\centering
	\caption{Stellar Properties for \starB. Systematic uncertainties are included on the values from \ariadne as in Table \ref{tab:stellar103524}.\\}
	\vspace{0.355cm}
	\begin{tabular}{lcc}
	\hline
	Property	&	Value		&Source\\
	\hline
	2MASS I.D.	& 03530331-3048164	&2MASS	\\
    Gaia source I.D. & 	4886825544715697792 &  \textit{Gaia DR2} \\
    TIC I.D. & TIC-166806344 & TIC8 \\
    \\
    \multicolumn{3}{l}{Astrometric Properties}\\
    R.A. &	\BRA &\textit{Gaia DR2}	\\
	Dec	& \BDec	&\textit{Gaia DR2}	\\
    $\mu_{{\rm R.A.}}$ (\masy) & \BpropRA & \textit{Gaia DR2} \\
	$\mu_{{\rm Dec.}}$ (\masy) & \BpropDec & \textit{Gaia DR2} \\
	Parallax (mas)             & \Bparallax & \textit{Gaia DR2} \\
    \\
    \multicolumn{3}{l}{Photometric Properties}\\
	V (mag)		&\BVmag 	&APASS\\
	B (mag)		&\BBmag		&APASS\\
	g (mag)		&\Bgmag		&APASS\\
	r (mag)		&\Brmag		&APASS\\
	i (mag)		&\Bimag		&APASS\\
    G (mag)		& \BGAIAmag	& \textit{Gaia DR2}\\
    J (mag)		&\BJmag	    &2MASS	\\
   	H (mag)		&\BHmag 	&2MASS	\\
	K (mag)		&\BKmag	&2MASS	\\
    W1 (mag)	&\BWmag		&WISE	\\
    W2 (mag)	&\BWWmag	&WISE	\\
    T (mag)     &\BTESSmag   &TIC8   \\
    \\
    \multicolumn{3}{l}{Derived Properties}\\

    Spectral type                           & \Bspectype    & \ariadne \\
    T$_{\rm eff}$ (K)                       & \Bteff        & \ispec\\
    \feh                            		& \Bmetal		& \ispec\\
    vsini (\kms)	                        & \Bvsini	    & \ispec\\
    $\gamma_{RV}$ (\kms)    & \Bgamma		& RV data \\
    log g                                   & \Blogg		& \ariadne\\
    \mstar (\msun)                          & \Bstarmass	& \ariadne\\
    \rstar (\rsun)                          & \Bstarradius	& \ariadne\\
    Age (Gyrs)			& \Bage				        	&SED fitting\\
    Distance (pc)	                        &  \Bdist	    & \ariadne\\
    $A_v$ (mag)                             & \BAv          & \ariadne \\
    
	\hline
    \multicolumn{3}{l}{2MASS \citep{2MASS}; APASS \citep{APASS};}\\
    \multicolumn{3}{l}{WISE \citep{WISE}; {\em Gaia} \citep{GAIA};}\\
    TIC8 \citep{Stassun19}\\
	\end{tabular}
    \label{tab:stellar103406}
\end{table}

\begin{table}
    \renewcommand{\arraystretch}{1.25}
	\centering
	\caption{Stellar Properties for \starC. Systematic uncertainties are included on the values from \ariadne as in Table \ref{tab:stellar103524}.}
	\begin{tabular}{lcc} 
	\hline
	Property	&	Value		&Source\\
	\hline
	2MASS I.D.	& 04513613-3413342	&2MASS	\\
    Gaia source I.D. & 	4873225513593736960 & \textit{Gaia DR2} \\
    TIC I.D. & TIC-1309019 & TIC8 \\
    \\
    \multicolumn{3}{l}{Astrometric Properties}\\
    R.A. &	\CRA    &\textit{Gaia DR2}	\\
	Dec	& \CDec	&\textit{Gaia DR2}	\\
    $\mu_{{\rm R.A.}}$ (\masy) & \CpropRA & \textit{Gaia DR2} \\
	$\mu_{{\rm Dec.}}$ (\masy) & \CpropDec & \textit{Gaia DR2} \\
	Parallax (mas)             & \Cparallax & \textit{Gaia DR2} \\
    \\
    \multicolumn{3}{l}{Photometric Properties}\\
	V (mag)		&\CVmag 	&APASS\\
	B (mag)		&\CBmag		&APASS\\
	g (mag)		&\Cgmag		&APASS\\
	r (mag)		&\Crmag		&APASS\\
	i (mag)		&\Cimag		&APASS\\
    G (mag)		&\CGAIAmag	& \textit{Gaia DR2}\\
    J (mag)		&\CJmag 	&2MASS	\\
   	H (mag)		&\CHmag 	&2MASS	\\
	K (mag)		&\CKmag 	&2MASS	\\
    W1 (mag)	&\CWmag		&WISE	\\
    W2 (mag)	&\CWWmag	&WISE	\\
    T (mag)     &\CTESSmag  &TIC8   \\
    \\
    \multicolumn{3}{l}{Derived Properties}\\

    Spectral type                           & \Cspectype    & \ariadne \\
    T$_{\rm eff}$ (K)                       & \Cteff        & \ispec\\
    \feh             		& \Cmetal		& \ispec\\
    vsini (\kms)	                        & \Cvsini	    & \ispec\\
    $\gamma_{RV}$ (\kms)    & \Cgamma		& RV data \\
    log g                                   & \Clogg		& \ariadne\\
    \mstar (\msun)                          & \Cstarmass	& \ariadne\\
    \rstar (\rsun)                          & \Cstarradius	& \ariadne\\
    Age (Gyrs)			& \Cage				        	&SED fitting\\
    Distance (pc)	                        &  \Cdist	    & \ariadne\\
    $A_v$ (mag)                             & \CAv     & \ariadne \\
    
	\hline
    \multicolumn{3}{l}{2MASS \citep{2MASS}; APASS \citep{APASS};}\\
    \multicolumn{3}{l}{WISE \citep{WISE}; {\em Gaia} \citep{GAIA};}\\
    TIC8 \citep{Stassun19}\\
	\end{tabular}
    \label{tab:stellar103867}
\end{table}

\begin{table}
    \renewcommand{\arraystretch}{1.25}
	\centering
	\caption{Stellar Properties for \starD. Systematic uncertainties are included on the values from \ariadne as in Table \ref{tab:stellar103524}.}
	\begin{tabular}{lcc}
	\hline
	Property	&	Value		&Source\\
	\hline
	2MASS I.D.	& 12021109-3532550	&2MASS	\\
    Gaia source I.D. & 	3462511310147530752 & \textit{Gaia DR2} \\
    TIC I.D. & TIC-142211778 & TIC8 \\
    \\
    \multicolumn{3}{l}{Astrometric Properties}\\
    R.A. &	\DRA	& \textit{Gaia DR2}	\\
	Dec	& \DDec	    & \textit{Gaia DR2}	\\
    $\mu_{{\rm R.A.}}$ (\masy) & \DpropRA & \textit{Gaia DR2}\\
	$\mu_{{\rm Dec.}}$ (\masy) & \DpropDec & \textit{Gaia DR2}\\
	Parallax (mas)             & \Dparallax  & \textit{Gaia DR2} \\
    \\
    \multicolumn{3}{l}{Photometric Properties}\\
	V (mag)		&\DVmag 	&APASS\\
	B (mag)		&\DBmag		&APASS\\
	g (mag)		&\Dgmag		&APASS\\
	r (mag)		&\Drmag		&APASS\\
	i (mag)		&\Dimag		&APASS\\
    G (mag)		&\DGAIAmag	&\textit{Gaia DR2}\\
    J (mag)		&\DJmag 	&2MASS	\\
   	H (mag)		&\DHmag 	&2MASS	\\
	K (mag)		&\DKmag	    &2MASS	\\
    W1 (mag)	&\DWmag		&WISE	\\
    W2 (mag)	&\DWWmag	&WISE	\\
    T (mag)     &\DTESSmag  &TIC8   \\
    \\
    \multicolumn{3}{l}{Derived Properties}\\

    Spectral type                           & \Dspectype    & \ariadne \\
    T$_{\rm eff}$ (K)                       & \Dteff        & \ispec\\
    \feh                            		& \Dmetal		& \ispec\\    vsini (\kms)	                        & \Dvsini	    & \ispec\\
    $\gamma_{RV}$ (\kms)   & \Dgamma		& RV data \\
    log g                                   & \Dlogg		& \ariadne\\
    \mstar (\msun)                          & \Dstarmass	& \ariadne\\
    \rstar (\rsun)                          & \Dstarradius	& \ariadne\\
    Age (Gyrs)			                    & \Dage			&SED fitting\\
    Distance (pc)	                        &  \Ddist	    & \ariadne\\
    $A_v$ (mag)                             & \DAv          & \ariadne \\

	\hline
    \multicolumn{3}{l}{2MASS \citep{2MASS}; APASS \citep{APASS};}\\
    \multicolumn{3}{l}{WISE \citep{WISE}; {\em Gaia} \citep{GAIA};}\\
    TIC8 \citep{Stassun19}\\
	\end{tabular}
    \label{tab:stellar104605}
\end{table}

%%%%%%%%%%%%%%%%%%%%%%%%%%%%%%%%%%%%%%%%%%%%%%%%%%%%%%%%%%%%%%%%%%%%%%%%%%%%%%%%%

\subsection{Global Modelling}
\label{sub:global}

We determined the physical parameters of the systems, including planet radius and mass, via a simultaneous fit to the photometric and spectroscopic data for each object. For this, we used the publicly available open-source astronomy software package \allesfitter \citep{gunther20, allesfittercode}, which unites the packages \ellc (light curve and RV models; \citealt{maxted16}), \emcee (MCMC sampling; \citealt{foremanmackey13}), and \celerite (Gaussian Process (GP) models; \citealt{foremanmackey17}). The combination of these packages allows \allesfitter to model a variety of signals, including multi-star systems, star spots, stellar variability, and transit-timing variations.

Initial MCMC fits required millions of steps to converge, so we chose the Nested Sampling approach to produce our global fits. Our priors for transit epoch and transit depth (\rpl/\rstar) were refined from \orion by generating quick MCMC fits to the NGTS data using \bruce, an open-source binary star and exoplanet analysis package\footnote{\url{https://github.com/samgill844/bruce}}. The results from the \ariadne SED fits were used as priors on the stellar parameters, specifically \rstar, \mstar\ and \teff, although we used the \teff error from \ispec as this is underestimated by \ariadne. Because all four host stars show no out-of-transit variability or trends, GP modelling was not necessary; indeed, an initial GP fit to the 2019 SAAO data of \starD\ incorrectly adjusted a data gap between the mid-transit and egress data points, resulting in a smaller transit depth than the true value. Furthermore, we accounted for instrumental offsets in the radial velocity data, as each object was observed with at least two different spectrographs. We adopted a quadratic limb-darkening law as parameterised in \citet{Kipping13}, and fit for the limb-darkening coefficients.

\medskip

We ran two fits for each planet: one in which the orbital eccentricity, $e$, was fixed at $0$, and one for which $e$ was allowed to vary freely. Each of the latter fits resulted in non-zero values of $e$, at $0.102\,^{+0.061}_{-0.068}$, $0.105\,^{+0.130}_{-0.079}$, $0.168\,^{+0.093}_{-0.101}$, and $0.035\,^{+0.033}_{-0.024}$ for \planetA, \planetB, \planetC, and \planetD, respectively. However, \citet{LucySweeney71} showed that many small values of $e$ are spurious, and define a probabilistic test to determine whether a small $e$ is statistically significant from 0. By adopting a 5\% level of significance, they find that if the condition $e > 2.45\sigma_e$ is satisfied (where $\sigma_e$ is the observational uncertainty on $e$), then one can be confident that the measured eccentricity is real. We find that all of our measurements fail to meet this criteria, and we therefore adopt the results from the global modelling fits in which $e$ is fixed at zero.

\medskip

We use \allesfitter to fit our data for the key physical parameters of each system, including the planet's orbital period P, the radius ratio \rpl/\rstar, and the radial velocity semi-amplitude K. Due to the large pixel scale of TESS (21" per pixel), the TESS FFI lightcurves for \starB\ and \starC\ contain additional faint (G $\gtrsim$ 17) neighbouring stars, and we therefore fit for a small dilution in the TESS data. We also note some blending for \starA\ in the NGTS data from a singular G $\sim19$ mag object (identified in \textit{Gaia DR2}), and account for this in our fitting. Note that dilution is defined in \allesfitter as:

\begin{equation}\label{dilution}
    D = 1 - \frac{F_{source}}{F_{source}+F_{blend}}
\end{equation}

\medskip

\noindent A full list of fitted properties and their values for each planet can be found in Table \ref{tab:planetparams}.

\medskip

Figure \ref{fig:allphotom} shows the phase-folded and single lightcurves for the NGTS, SAAO, and TESS data for each object, where the red lines indicate 20 models generated from randomly drawn posterior samples of the \allesfitter fits. Similarly, the phase-folded radial velocity data for each star is shown in Figure \ref{fig:allRVs}, with 20 generated radial velocity models.

\begin{table*}
    \begingroup
    \renewcommand{\arraystretch}{1.3}
    \centering
    \caption{Planetary properties for each system from \allesfitter (see Equation \ref{dilution} for a definition of dilution, D)}
    \label{tab:planetparams}
    \medskip
    \begin{subtable}{.45\linewidth}
        \caption{Planetary Properties for \planetA}
        \smallskip
        \centering
	    \begin{tabular}{lc}
	        \hline
	        Property	        &	Value       \\
	        \hline
            P (days)		    &	\Abperiod   \\
 	        T$_C$ (BJD)		    &	\Abtc	    \\
            T$_{14}$ (hours)    &   \Abduration \\
            $R_p/R_{*}$         &   \Abrratio   \\
            $a/R_{*}$	    	&    \Abaoverr   \\
            $b$                 &   \Abimpact   \\
	        K (\ms) 	        &   \Akamp	    \\
            e 			        &   0 (fixed)  	\\
            \mpl (\mjup)        &   \Abmass	    \\
            \rpl (\rjup)        &   \Abradius   \\
            $\rho_{p}$ (\gccc)  &   \Abdensity  \\
            a (AU)              &   \Abau       \\
            T$_{eq}$ (K) $^{\hyperlink{Teq}{\dagger}}$        &   \AbTeq  	\\
  	        Irradiation (Wm$^{-2}$)      &   \Abflux 	\\
  	        D$_{\mathrm{NGTS}}$ &   \ANGTSdil   \\
	        \hline
    	\end{tabular}
    \end{subtable}%
    \begin{subtable}{.45\linewidth}
        \caption{Planetary Properties for \planetB}
        \smallskip
        \centering
	    \begin{tabular}{lc}
	        \hline
	        Property	        &	Value       \\
	        \hline
            P (days)		    &	\Bbperiod   \\
 	        T$_C$ (BJD)		    &	\Bbtc	    \\
            T$_{14}$ (hours)    &   \Bbduration \\
            $R_p/R_{*}$         &   \Bbrratio   \\
            $a/R_{*}$	    	&   \Bbaoverr   \\
            $b$                 &   \Bbimpact   \\
	        K (\ms) 	        &   \Bkamp	    \\
            e 			        &   0 (fixed)  	\\
            \mpl (\mjup)        &   \Bbmass	    \\
            \rpl (\rjup)        &   \Bbradius   \\
            $\rho_{p}$ (\gccc)  &   \Bbdensity  \\
            a (AU)              &   \Bbau       \\
            T$_{eq}$ (K) $^{\hyperlink{Teq}{\dagger}}$       &   \BbTeq  	\\
  	        Irradiation (Wm$^{-2}$)      &   \Bbflux 	\\
  	        D$_{\mathrm{TESS}}$ &   \BTESSdil   \\
	        \hline
        \end{tabular}
    \end{subtable}
    
    \bigskip
    
    \medskip
    
    \begin{subtable}{.45\linewidth}
        \caption{Planetary Properties for \planetC}
        \smallskip
        \centering
	    \begin{tabular}{lc}
	        \hline
	        Property	        &	Value       \\
	        \hline
            P (days)		    &	\Cbperiod   \\
 	        T$_C$ (BJD)		    &	\Cbtc	    \\
            T$_{14}$ (hours)    &   \Cbduration \\
            $R_p/R_{*}$         &   \Cbrratio   \\
            $a/R_{*}$	    	&   \Cbaoverr   \\
            $b$                 &   \Cbimpact   \\
	        K (\ms) 	        &   \Ckamp	    \\
            e 			        &   0 (fixed)  	\\
            \mpl (\mjup)        &   \Cbmass	    \\
            \rpl (\rjup)        &   \Cbradius   \\
            $\rho_{p}$ (\gccc)  &   \Cbdensity  \\
            a (AU)              &   \Cbau       \\
            T$_{eq}$ (K) $^{\hyperlink{Teq}{\dagger}}$       &   \CbTeq  	\\
   	        Irradiation (Wm$^{-2}$)      &   \Cbflux 	\\
  	        D$_{\mathrm{TESS}}$ &   \CTESSdil   \\
	        \hline
        \end{tabular}
    \end{subtable}%
    \begin{subtable}{.45\linewidth}
        \caption{Planetary Properties for \planetD}
        \smallskip
        \centering
	    \begin{tabular}{lc}
	        \hline
	        Property	        &	Value       \\
	        \hline
            P (days)		    &	\Dbperiod   \\
	        T$_C$ (BJD)		    &	\Dbtc	    \\
            T$_{14}$ (hours)    &   \Dbduration \\
            $R_p/R_{*}$         &   \Dbrratio   \\
            $a/R_{*}$	    	&   \Dbaoverr   \\
            $b$                 &   \Dbimpact   \\
	        K (\ms) 	        &   \Dkamp	    \\
            e 			        &   0 (fixed)  	\\
            \mpl (\mjup)        &   \Dbmass	    \\
            \rpl (\rjup)        &   \Dbradius   \\
            $\rho_{p}$ (\gccc)  &   \Dbdensity  \\
            a (AU)              &   \Dbau       \\
            T$_{eq}$ (K) $^{\hyperlink{Teq}{\dagger}}$       &   \DbTeq  	\\
 	        Irradiation (Wm$^{-2}$)      &   \Dbflux 	\\
	        \hline
	        \hspace{0.5cm}
        \end{tabular}
    \end{subtable}%
    \\\vspace{0.3cm}
    \endgroup
    %\begin{flushleft}
    {$^{\dagger}\hypertarget{Teq}$ T$_{eq} =$ T$_{\mathrm{eff; }s} \cdot \frac{(1-A)^{1/4}}{E} \cdot \sqrt{\frac{R_{s}}{2a}}$, where albedo $A = 0.3$ and emissivity $E = 1$ \hspace{4.5cm}}
    %\end{flushleft}
\end{table*}

%%%%%%%%%%%%%%%%%%%%%%%%%%%%%%%%%%%%%%%%%%%%%%%%%%%%%%%%%%%%%%%%%%%%%%%%%%%%%%%%%

\section{Inflation}
\label{sec:Inflation}

\begin{table*}
	\centering
	\caption{Quantifying the inflation of \planetA, \planetB, \planetC, and \planetD. $R_{\mathrm{observed}}$ describes the radius derived from global fits to the observational data, whilst $R_{\mathrm{non-inflated}}$ and $R_{\mathrm{inflated}}$ describe the predicted radius from non-inflationary evolutionary models and inflationary forward models, respectively.}
	\label{tab:inflation}
	\begin{tabular}{cccccccc}
    \hline
    \hline
    Target & Irradiation & Mass (\mjup) & $\Delta R$ (\rjup)* & $R_{\mathrm{inflated}}$ (\rjup)* & $R_{\mathrm{non-inflated}}$ (\rjup)** & $R_{\mathrm{observed}}$ (\rjup)\\
\hline
\hline
\planetA & \Abflux  & \Abmass & $0.18\pm0.07$ & $1.16 \pm 0.08$   & $1.02-1.45$ & \Abradius \\
\hline
\planetB & \Bbflux  & \Bbmass & $0.22\pm0.07$ & $1.20 \pm 0.08$   & $1.01-1.07$ & \Bbradius \\
\hline
\planetC & \Cbflux  & \Cbmass & $0.49\pm0.07$ & $1.47 \pm 0.08$   & $1.01-1.08$ & \Cbradius \\
\hline
\planetD & \Dbflux  & \Dbmass & $0.40\pm0.10$ & $1.38 \pm 0.11$   & $0.99-1.06$ &\Dbradius \\
	\hline
	\multicolumn{7}{l}{*From \citet{Sestovic18}; **From \citet{Baraffe08}}\\
    \end{tabular}
\end{table*}

All four planets presented in this paper have sub-Jovian masses and super-Jovian radii. Most notably, the radius of \planetD\ is \Dbradius\,\rjup, but with a mass of only \Dbmass\,\mjup. This is not unusual for close-in gaseous planets: the region of parameter space pertaining to low density hot Jupiters is well populated, and is driven by inflation mechanisms which correlate with stellar irradiation \citep{Laughlin11, Weiss13, Thorngren18}.

Studies by both \citet{demory11} and \citet{Miller11} suggest that these mechanisms become effective above an incident flux of $\sim$\inflationflux. By using the stellar luminosities and orbital parameters of each system, we calculated the irradiation received by each planet and found that all four are irradiated above this threshold (see Table~\ref{tab:planetparams} for values). As such, it is reasonable to expect that at least some of these planets may be affected by inflation processes and therefore exhibit radii which are larger than predicted.

To parameterise this, we follow the procedure outlined in \citet{Costes20}, which utilises the work of \citet{Baraffe08} to compare planetary radii with predictions from evolutionary models, and the work of \citet{Sestovic18} to discern the predicted additional radius change from inflation, $\Delta R$.

\medskip

\citet{Baraffe08} (hereafter B08) produced theoretical planetary evolution models which account for uncertainties in previous models, most notably quantifying the impact of heavy element enrichment. Their results include tables of predicted planetary radii which vary with the system's physical parameters. Using these tables and adjusting for spectral type, we calculated the expected radii for all four planets and compared the result with the radii derived from our global fits. A discrepancy between the two values could indicate that inflation mechanisms have affected the radius. As B08 account for a decrease in radius with increasing heavy element mass, we assume a mass fraction of heavy material of $0.02 - 0.1$ for each planet, which provides a realistic upper limit for the non-inflated radii.

\citet{Sestovic18} (hereafter S18) implement hierarchical Bayesian modelling and a forward model to infer relationships between incident flux, radius, and mass using a population of 286 hot Jupiters with measured radii and masses. The resulting relations for $\Delta R$ vary according to four different mass regimes: below 0.37\,\mjup, between $0.37-0.98$\,\mjup, between $0.98-2.50$\,\mjup and over 2.50\,\mjup. In this case, all four of the planets fall into the same mass range ($0.37-0.98$\,\mjup), and we therefore use the corresponding equation to derive the expected radius increase due to inflation for each:

\begin{equation} \label{Sestovic_model}
    \Delta R = 0.70 \cdot (\log_{10}F-5.5), \hspace{10pt} 0.37\leq \textstyle\frac{M}{M_J}<0.98
\end{equation}

\noindent (see S18 Equation 11). Note that, in the S18 models, the `baseline' radius, $R$, is set to $0.98\pm0.04$\,\rjup\ for this mass regime. We therefore calculate the total expected inflated radius from the models of S18 as

\begin{equation} \label{Sestovic_baseline}
    R_{\mathrm{Inflated}} = 0.98 + \Delta R
\end{equation}

\noindent and compare this to the `true' radii of the planets derived from global modelling. 

\section{Discussion}
\label{sec:Discussion}

We present our inflation results in Table~\ref{tab:inflation}, and summarise below:
\\

\noindent \textbf{\planetA:} Due to the poorly constrained age for this planet, the range of possible values for $R_{\mathrm{non-inflated}}$ is so broad that it encompasses and exceeds the expected $R_{\mathrm{inflated}}$ values. As such, $R_{\mathrm{observed}}$ and its uncertainties appear consistent with both inflated and non-inflated radii, and we are therefore unable to draw firm conclusions about the nature of inflation of \planetA. \\

\noindent \textbf{\planetB:} $R_{\mathrm{observed}}$ does not agree to within uncertainties with $R_{\mathrm{non-inflated}}$, but it is consistent with $R_{\mathrm{inflated}}$. However, the discrepancy between $R_{\mathrm{observed}}$ and $R_{\mathrm{non-inflated}}$ is less than $2 \sigma$, so the possibility that this planet is not inflated is also non-negligable. \\

\noindent \textbf{\planetC:} We find that, whilst $R_{\mathrm{observed}}$ is higher than $R_{\mathrm{non-inflated}}$, it is not as large as $R_{\mathrm{inflated}}$. Although the discrepancy between $R_{\mathrm{observed}}$ and $R_{\mathrm{non-inflated}}$ is less than $2\sigma$, note that, due to the high irradiation of \planetC, the planet exists in a region of parameter space in which S18 find that there is no evidence for a population of non-inflated hot Jupiters. It is therefore probable that this planet is inflated and that the models from S18 fail to predict the true radius value. \\

\noindent \textbf{\planetD:} For this planet, the large uncertainties of $R_{\mathrm{observed}}$ mean that \planetD\ can be described by both $R_{\mathrm{inflated}}$ and $R_{\mathrm{non-inflated}}$; although as with \planetC, we note that the large incident flux implies that this planet is likely to be inflated. In addition, we find that the lower boundary of $R_{\mathrm{observed}}$ falls into S18's low-mass regime (below 0.37 $M_J$), and we therefore use the corresponding equation for $\Delta R$ from S18 to derive an alternative $R_{\mathrm{inflated}}$ value of $0.96 \pm 0.9$. In this case, again $R_{\mathrm{observed}}$ is in agreement with $R_{\mathrm{inflated}}$ to within errors, but now $R_{\mathrm{inflated}}$ and $R_{\mathrm{non-inflated}}$ are also entirely consistent with one another, leading us inclined to disregard this result.

\begin{figure}
    \centering
    \includegraphics[width=0.5\textwidth]{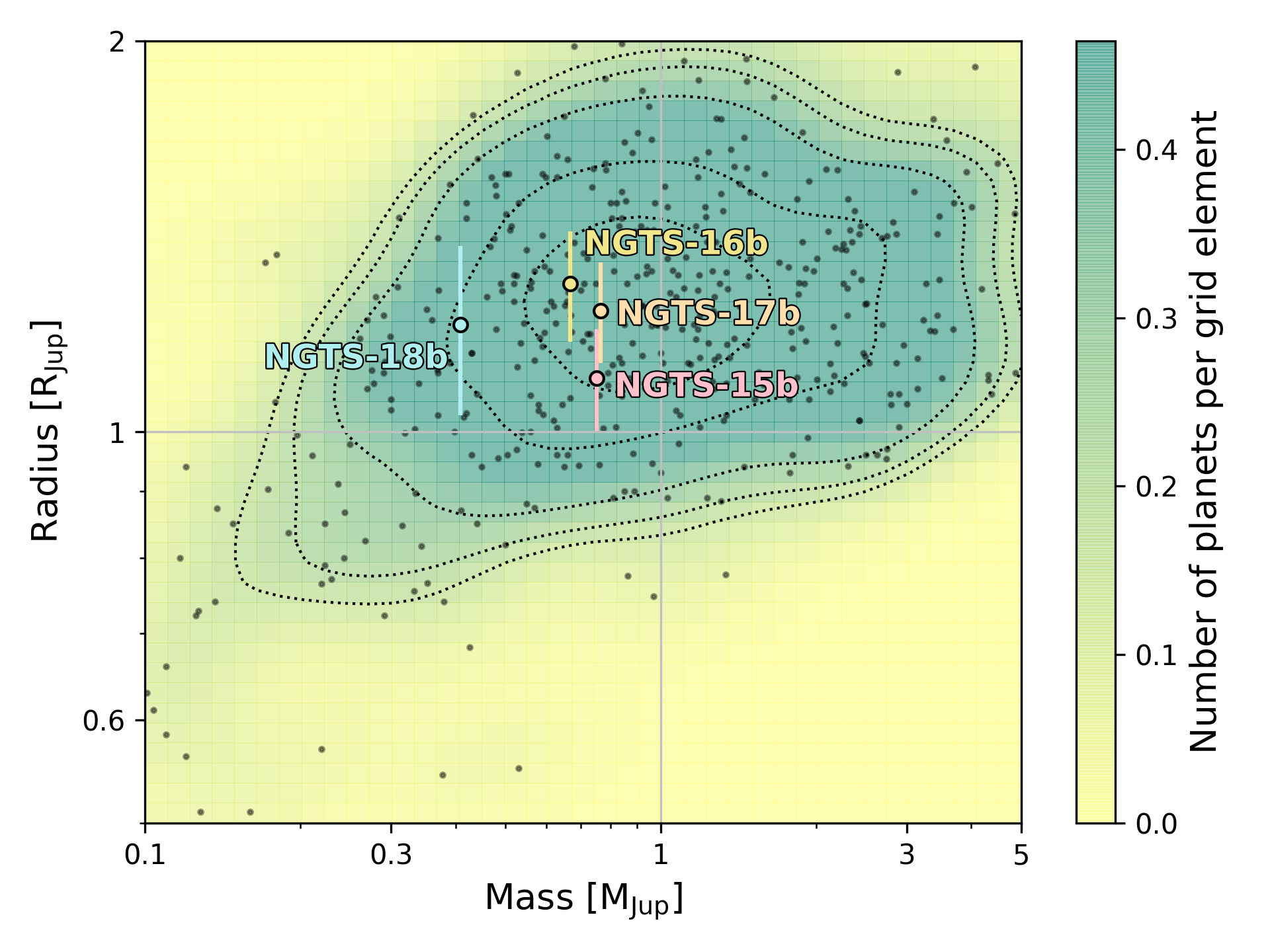}
    \caption{Exoplanets with confirmed masses and radii (grey points) from the NASA Exoplanet Archive (\url{https://exoplanetarchive.ipac.caltech.edu/}). The background and the dotted black contour lines highlight the point density per grid element of our sample. The hot Jupiters presented in this paper are plotted with error bars, and can be seen to lie comfortably within the general population of hot Jupiters.  \planetA\ to \planetD\ are labeled with coloured circles on the plot with associated uncertainties.}
    \label{fig:mass_radius}
\end{figure}

\begin{figure}
    \centering
    \includegraphics[width=0.5\textwidth]{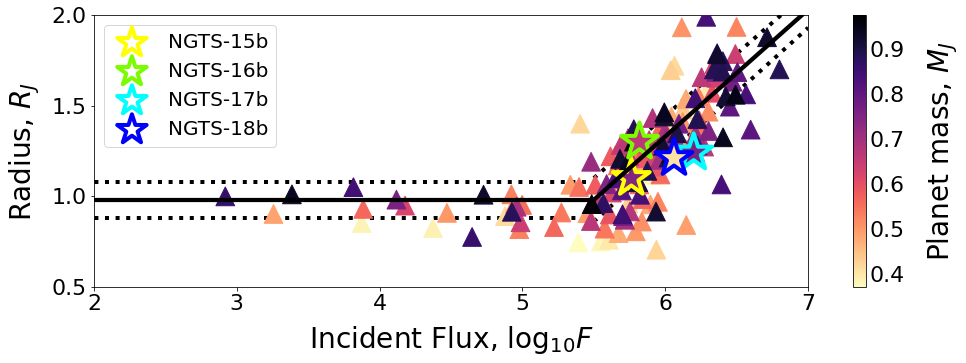}
    \caption{Exoplanets from the Exoplanet Orbit Database \citep{exo_orbit_database_ref} with masses between $0.37-0.97$ \mjup. We overplot the model (solid black line) from \citet{Sestovic18} with the associated standard deviation in $R$ (dotted black line) for hot Jupiter inflation in this mass regime. The hot Jupiters presented in this paper are plotted as stars with coloured outlines.}
    \label{fig:inflation}
\end{figure}

\medskip

Although all of the incident fluxes are larger than the typically accepted threshold for inflation ($2\times10^{5}$\,Wm$^{-2}$) \citep{demory11, Miller11}, it is possible that some of these planets, particularly \planetA, are not inflated. However, this boundary only describes the minimum limit at which planets begin to appear inflated, and so this result is not unexpected. In addition, S18 finds that above $\sim10^{6}$\,Wm$^{-2}$, there is no evidence for a population of non-inflated hot Jupiters in the mass range $0.37-0.98$\,\mjup. The results presented here do not necessarily support this claim, but as the two planets experiencing irradiation in excess of $\sim10^{6}$\,Wm$^{-2}$ do show some evidence for inflation, we take the results from S18 as an indication that \planetC\ and \planetD\ are more likely to be inflated than not.

\medskip

It is interesting that, based on the models of S18, we expect both \planetA\ and \planetB\ to possess a similarly inflated radius. Whilst it is possible that both planets do correspond with the predicted inflated radii from S18, it is also probable that \planetA\ is not inflated at all, especially if the system is young. However, we note that, whilst S18's model successfully describes the general shape of the hot Jupiter population in the radius-flux plane (see Figure \ref{fig:inflation}), the variation in latent parameters such as heavy element fraction and system age is only accounted for in the scatter of the model; currently, the models are solely dependant on incident flux (within mass bins). We speculate that the variation in additional latent factors for the planets presented here would provide an explanation as to why two similarly irradiated planets may exhibit different levels of radius inflation. As previously mentioned, the age of \planetA\ could reasonably lead to an uninflated radius; it is already known that hot Jupiter radii vary with age \citep{Baraffe08, Miller09, Thorngren18}, and new evidence is emerging which suggests that gaseous planets may `reinflate' at late times \citep{Hartman16, Komacek2020, Thorngren2020, Thorngren2021}. In addition, differing system ages could indicate a difference in planetary migration times onto short orbits, and there may have been a variation in the level of stellar irradiation since that time. This may be one explanation for the systems presented here, as the SED ages, while uncertain for the younger system, would appear to align with this. Furthermore, as noted previously, the fraction of heavy elements present in a planet will influence the radius inflation of hot Jupiters, with radii expected to decrease as $Z$ increases \citep{Thorngren16, Thorngren2021}.

Whilst these parameters are described indirectly by the models of S18 as the intrinsic physical scatter in the data, the detailed effects are as-of-yet unidentified. With the number of hot Jupiter candidates increasing more than two fold since the construction of the sample from S18, it would be pertinent to update these models with a view to understanding whether the inclusion of additional hyperparameters in the Bayesian model may describe the hot Jupiter inflation relationships with greater accuracy.

\medskip

Finally, despite the varying response to inflation mechanisms, we find that all four planets lie comfortably amongst the general population of hot Jupiters in the mass-radius plane (see Figure \ref{fig:mass_radius}). Additionally, all four planets again occupy a densely-populated region of radius-flux parameter space (see Figure \ref{fig:inflation}).

\section{Conclusions}

We report the discovery of four hot Jupiter planets: \planetA, \planetB, \planetC, and \planetD. Each planet was originally identified from photometry from the Next Generation Transit Survey (NGTS), and was confirmed through follow-up observations carried out at the South African Astronomical Observatory (SAAO) with the 1.0\,m and Lesedi telescopes, and radial velocity measurements made with the HARPS, CORALIE and FEROS spectrographs. Additional photometry from the Transiting Exoplanet Survey Satellite (TESS) was also acquired for three of the four targets.

Global fits to these data were produced using the open-source astronomy software package \allesfitter \citep{gunther20,allesfittercode}, and yielded masses, radii, and orbital periods consistent with hot Jupiter planets (see Table \ref{tab:planetparams}). Spectral analysis via \ispec \citep{Blanco14, blanco19} and SED fitting via \ariadne \citep{Vines21} revealed the properties of the host stars in each system, which were all found to be main-sequence G-type stars (see Tables \ref{tab:stellar103524}, \ref{tab:stellar103406}, \ref{tab:stellar103867}, and \ref{tab:stellar104605}).

As part of our analysis, we noted that all four planets received a level of irradiation that surpassed the expected threshold for the onset of planetary inflation mechanisms \citep{demory11, Miller11}. As such, we sought to characterise any potential inflation by comparing our derived radii with predictions from evolutionary models \citep{Baraffe08}. In addition, we examined the predicted additional radius change due to inflation, $\Delta R$, through the flux-mass-radius relations outlined in \citet{Sestovic18}. We found that two of the planets are likely inflated when compared with non-inflationary models, and it is reasonable to assume that \planetD, although consistent with both inflated and non-inflated solutions, is also inflated due to its high incident flux. We were unable to draw any firm conclusions on the nature of inflation for \planetA\ due to the poorly constrained age for this system resulting in a broad range of non-inflated radius values. Furthermore, we note some disparity between the radii derived from global modelling and those predicted by the inflationary forward model of \citet{Sestovic18}, although all four planets were found to fit within the general hot Jupiter population (see Figures~\ref{fig:mass_radius}~\&~\ref{fig:inflation}). We suggest that the inclusion of both new hot Jupiter data and additional hyperparameters which describe latent parameters, such as system age, into the Bayesian model may refine the relations further.

\section*{Acknowledgements}
Based on data collected under the NGTS project at the ESO La Silla Paranal Observatory.  The NGTS facility is operated by the consortium institutes with support from the UK Science and Technology Facilities Council (STFC)  project ST/M001962/1. RHT is supported by an STFC studentship.

This study is based on observations collected at the European Southern Observatory under ESO programmes 0103.C-0719 and 0104.C-0588 (P.I. Bouchy).

This paper uses observations made at the South African Astronomical Observatory (SAAO).

This paper includes data collected by the TESS mission. Funding for the TESS mission is provided by the NASA Explorer Program. This research has made use of the NASA Exoplanet Archive, which is operated by the California Institute of Technology, under contract with the National Aeronautics and Space Administration under the Exoplanet Exploration Program.

This research has made use of the Exoplanet Orbit Database and the Exoplanet Data Explorer at \url{https://www.exoplanets.org}.

This publication makes use of The Data \& Analysis Center for Exoplanets (DACE), which is a facility based at the University of Geneva (CH) dedicated to extrasolar planets data visualisation, exchange and analysis. DACE is a platform of the Swiss National Centre of Competence in Research (NCCR) PlanetS, federating the Swiss expertise in Exoplanet research. The DACE platform is available at \url{https://dace.unige.ch}.

This work has made use of data from the European Space Agency (ESA) mission
{\it Gaia} (\url{https://www.cosmos.esa.int/gaia}), processed by the {\it Gaia}
Data Processing and Analysis Consortium (DPAC,
\url{https://www.cosmos.esa.int/web/gaia/dpac/consortium}). Funding for the DPAC
has been provided by national institutions, in particular the institutions
participating in the {\it Gaia} Multilateral Agreement.

The research leading to these results has received funding from the European Research Council under the European Union's Seventh Framework Programme (FP/2007-2013) / ERC Grant Agreement n. 320964 (WDTracer).

Contributions by authors from the University of Warwick were supported by STFC consolidated grants ST/P000495/1 and ST/T000406/1.

EG gratefully acknowledges support from the David and Claudia Harding Foundation in the form of a Winton Exoplanet Fellowship.

JSJ acknowledges support by FONDECYT grant 1201371, and partial support from CONICYT project Basal AFB-170002.

ACh and PhE acknowledge the support of the DFG priority program SPP 1992 ``Exploring the Diversity of Extrasolar Planets" (RA 714/13-1)

JIV acknowledges support of CONICYT-PFCHA/Doctorado Nacional-21191829.

MNG acknowledges support from the MIT Kavli Institute as a Juan Carlos Torres Fellow.

We thank the Swiss National Science Foundation  (SNSF) and the Geneva University for their continuous support to our planet search programs. This work has been in particular carried out in the frame of the National Centre for Competence in Research ``PlanetS" supported by SNSF.

\section*{Data Availability}

The data underlying this article will be shared on reasonable request to the corresponding author.

%%%%%%%%%%%%%%%%%%%%%%%%%%%%%%%%%%%%%%%%%%%%%%%%%%

%%%%%%%%%%%%%%%%%%%% REFERENCES %%%%%%%%%%%%%%%%%%

\bibliographystyle{mnras}
%\bibliography{mybibliography}

%%%%%%%%%%%%%%%%%%%%%%%%%%%%%%%%%%%%%%%%%%%%%%%%%%

%%%%%%%%%%%%%%%%% APPENDICES %%%%%%%%%%%%%%%%%%%%%

\appendix
\section{Radial velocity data}
\label{sec:appendix}

\begin{table*}
	\centering
	\caption{Summary of radial velocity data from HARPS, CORALIE and FEROS}
	\label{tab:rvfullsummary}
	\begin{tabular}{cccccccc} 
	\hline
	\hline
Target & BJD			&	RV		& RV err &	FWHM & 	contrast & BIS & Instrument\\
& (-2450000)	& (\kms)& (\kms)&(\kms) & &(\kms)& \\
		\hline
		\hline
& 8423.852 & 34.764 & 0.102 & 8.204 & 75.167 & 0.421    & CORALIE \\
& 8428.738 & 34.574 & 0.100 & 8.117 & 68.305 & 0.013    & CORALIE \\
& 8493.712 & 34.632 & 0.157 & 8.681 & 89.838 & -0.418   & CORALIE \\
& 8497.732 & 34.561 & 0.123 & 8.252 & 75.461 & 0.146    & CORALIE \\
& 8528.658 & 34.722 & 0.132 & 7.869 & 75.167 & 0.421    & CORALIE \\
& 8538.571 & 34.613 & 0.143 & 8.466 & 72.993 & -1.919   & CORALIE \\
& 8543.582 & 34.545 & 0.137 & 8.398 & 60.515 & 0.073    & CORALIE \\
& 8736.864 & 34.601 & 0.018 & 10.101& 0.660  & -0.052   & FEROS\\
& 8739.802 & 34.549 & 0.022 & 6.683 & 30.008 & 0.118    & HARPS \\
& 8739.858 & 34.595 & 0.024 & 9.706 & 0.690	 & -0.134   & FEROS\\
\starA & 8740.862 & 34.803 & 0.029 & 9.703 & 0.710	& -0.037 & FEROS \\
& 8741.797 & 34.786 & 0.017 & 6.687 & 31.418 & 0.014    & HARPS \\
& 8741.833 & 34.799 & 0.018 & 9.685 & 0.680	 & -0.033	& FEROS \\
& 8742.782 & 34.696 & 0.024 & 9.440 & 0.690  & -0.220   & FEROS \\
& 8742.863 & 34.625 & 0.019 & 9.705 & 0.670	 & -0.060   & FEROS \\
& 8744.850 & 34.856 & 0.021 & 9.769 & 0.670  & -0.060	& FEROS \\
& 8782.766 & 34.581 & 0.031 & 6.959 & 32.833 & 0.013    & HARPS \\
& 8783.767 & 34.683 & 0.023 & 6.857 & 33.327 & 0.055    & HARPS \\
& 8784.793 & 34.741 & 0.027 & 6.850 & 33.002 & 0.067    & HARPS \\
& 8808.728 & 34.608 & 0.020 & 6.833 & 33.754 & -0.079   & HARPS \\
& 8820.798 & 34.733 & 0.010 & 6.930 & 34.717 & 0.013    & HARPS \\

\hline

& 8847.635 & 29.126 & 0.014 & 9.970  & 0.600 & 0.063  & FEROS \\
& 8848.650 & 29.078 & 0.016 & 10.217 & 0.600 & -0.075 & FEROS \\
& 8849.653 & 29.079 & 0.015 & 10.058 & 0.590 & 0.035  & FEROS \\
& 8850.616 & 29.209 & 0.014 & 10.002 & 0.600 & -0.012 & FEROS \\
& 8851.619 & 29.219 & 0.014 & 10.034 & 0.600 & -0.003 & FEROS \\
& 8852.614 & 29.072 & 0.015 & 10.112 & 0.600 & -0.063 & FEROS \\
\starB & 8869.629 & 29.155 & 0.009 & 7.650 & 53.996 & -0.026 & HARPS \\
& 8875.604 & 29.190 & 0.009 & 7.697 & 53.738 & -0.026 & HARPS \\
& 8886.579 & 29.136 & 0.011 & 7.693 & 51.044 & -0.040 & HARPS \\
& 8887.572 & 29.058 & 0.010 & 7.649 & 51.040 & -0.032 & HARPS \\
& 8927.511 & 29.148 & 0.009 & 7.625 & 53.670 & -0.039 & HARPS \\
& 8930.509 & 29.090 & 0.030 & 7.605 & 51.292 & -0.003 & HARPS \\

\hline

& 8737.879 & 34.842 & 0.019 & 10.145 & 0.670  & -0.113 & FEROS \\
& 8739.818 & 34.894 & 0.021 & 10.094 & 0.690  & -0.039 & FEROS \\
& 8741.857 & 34.763 & 0.016 & 9.878  & 0.680  & 0.032  & FEROS \\
& 8742.759 & 34.910 & 0.023 & 10.090 & 0.700  & 0.045  & FEROS \\
& 8744.784 & 34.768 & 0.024 & 10.102 & 0.690  & -0.032 & FEROS \\
& 8745.853 & 35.032 & 0.039 & 10.180 & 0.700  & 0.138  & FEROS \\
& 8757.770 & 34.699 & 0.160 & 8.641  & 46.025 & 0.439  & CORALIE \\
& 8791.689 & 34.813 & 0.147 & 8.823  & 44.152 & 0.130  & CORALIE \\
& 8804.645 & 34.805 & 0.163 & 8.977  & 50.196 & 0.296  & CORLAIE \\
& 8847.672 & 34.801 & 0.017 & 10.239 & 0.660  & -0.037 & FEROS \\
\starC & 8848.674 & 34.702 & 0.017 & 9.955 & 0.660 & 0.051 & FEROS \\
& 8849.677 & 34.945 & 0.019 & 10.190 & 0.660  & -0.024 & FEROS \\
& 8850.692 & 34.925 & 0.016 & 10.165 & 0.670  & -0.016 & FEROS \\
& 8851.662 & 34.742 & 0.016 & 10.145 & 0.670  & -0.019 & FEROS \\
& 8852.669 & 34.769 & 0.015 & 10.187 & 0.670  & -0.030 & FEROS \\
& 8861.678 & 34.573 & 0.176 & 8.926  & 45.527 & -0.231 & CORALIE \\
& 8864.664 & 34.672 & 0.191 & 8.971  & 48.217 & 0.255  & CORALIE \\
& 8869.639 & 34.809 & 0.174 & 8.683  & 46.440 & -0.327 & CORALIE \\
& 8882.644 & 34.981 & 0.184 & 8.650  & 48.722 & 0.467  & CORALIE \\
& 8893.664 & 34.694 & 0.172 & 8.465  & 47.912 & 0.122  & CORALIE \\

		\hline
	\end{tabular}
\end{table*}

\begin{table*}
	\centering
	\caption{Summary of radial velocity data from HARPS, CORALIE and FEROS (cont.)}
	\label{tab:rvs103524}
	\begin{tabular}{cccccccc}
	\hline
	\hline
Target & BJD			&	RV		&RV err &	FWHM& 	contrast & BIS & Instrument\\
& (-2450000)	& (\kms)& (\kms)&(\kms) & &(\kms)& \\
		\hline
		\hline

& 8624.587 & 5.204 & 0.171 & 8.523 & 54.102 & -0.056 & CORALIE \\
& 8847.770 & 5.207 & 0.017 & 9.177 & 0.620  & 0.034  & FEROS \\
& 8848.777 & 5.151 & 0.018 & 9.912 & 0.650  & 0.067  & FEROS \\
& 8849.800 & 5.171 & 0.019 & 9.321 & 0.630  & -0.067 & FEROS \\
& 8850.817 & 5.340 & 0.018 & 9.872 & 0.650  & 0.056  & FEROS \\
& 8852.793 & 5.207 & 0.016 & 9.746 & 0.650  & 0.053  & FEROS \\
& 8862.770 & 5.173 & 0.160 & 7.402 & 61.367 & -0.494 & CORALIE \\
& 8910.839 & 5.182 & 0.010 & 7.556 & 46.878 & -0.038 & HARPS \\
\starD & 8911.815 & 5.206 & 0.009 & 7.510 & 47.123 & -0.033 & HARPS \\
& 8912.822 & 5.106 & 0.008 & 7.510 & 47.168 & -0.001 & HARPS \\
& 8916.780 & 5.150 & 0.013 & 7.438 & 46.887 & -0.003 & HARPS \\
& 8918.843 & 5.130 & 0.014 & 7.557 & 45.336 & 0.048 & HARPS \\
& 8925.810 & 5.165 & 0.012 & 7.536 & 47.105 & 0.028 & HARPS \\
& 8926.836 & 5.244 & 0.011 & 7.510 & 46.404 & -0.021 & HARPS \\
& 8928.814 & 5.137 & 0.012 & 7.574 & 46.375 & -0.035 & HARPS \\
& 8929.846 & 5.235 & 0.013 & 7.567 & 46.364 & 0.038 & HARPS \\
& 8931.808 & 5.131 & 0.015 & 7.535 & 46.549 & 0.006 & HARPS \\

		\hline
	\end{tabular}
\end{table*}

%%%%%%%%%%%%%%%%%%%%%%%%%%%%%%%%%%%%%%%%%%%%%%%%%%

% Don't change these lines
\bsp	% typesetting comment
\label{lastpage}
\end{document}